%% file: artss15_arxiv.tex
\documentclass{aa}
\usepackage[utf8]{inputenc}
\usepackage[colorlinks]{hyperref}
\usepackage{amsmath}
\usepackage{amssymb}
\usepackage{microtype}
\usepackage[T1]{fontenc}
\usepackage{lmodern}
\usepackage{natbib}
\usepackage{graphicx}
\usepackage{microtype}
\usepackage{color}
\usepackage[dvipsnames]{xcolor}
\usepackage{multirow}
\usepackage{xspace}
\usepackage[normalem]{ulem}
\usepackage{sidecap}

\usepackage{lscape}

\usepackage{pdflscape} 

\bibpunct{(}{)}{;}{a}{}{,} 

\def\aap{A\&A}

\def\apjs{ApJS}

\def\apjl{ApJLett}

\def\srg{\textit{SRG}}
\def\art{ART-XC}
\def\erosita{eROSITA}
\def\uhuru{\textit{Uhuru}}

\def\maxi{MAXI}

\def\swift{\textit{Swift}}

\def\xmm{\textit{XMM-Newton}}
\def\fermi{\textit{Fermi}}
\def\gaia{\textit{Gaia}}

\def\ass{ARTSS1-5}

\def\arcmin{\hbox{$^\prime$}}
\def\arcsec{\hbox{$^{\prime\prime}$}}
\def\flux{erg\,s$^{-1}$\,cm$^{-2}$}

\def\fsp{f_{\rm sp}}

\begin{document}

\title{SRG/\art\ all-sky X-ray survey: Catalog of sources detected during the first five surveys\thanks{The catalog is only available in electronic form at the CDS via anonymous ftp to cdsarc.u-strasbg.fr (130.79.128.5) or via \url{http://cdsweb.u-strasbg.fr/cgi-bin/qcat?J/A+A/} and at \url{http://srg.cosmos.ru}.}
}

\author{S. Sazonov\inst{1}\thanks{E-mail: sazonov@cosmos.ru} \and R. Burenin\inst{1} \and  E. Filippova\inst{1} \and R.~Krivonos\inst{1} \and V. Arefiev\inst{1} \and K. Borisov\inst{2}\and M. Buntov\inst{1} \and C.-T. Chen\inst{3} \and S. Ehlert\inst{5} \and S. Garanin\inst{4} \and M. Garin\inst{4} \and S. Grigorovich\inst{4} \and I. Lapshov\inst{1} \and V. Levin\inst{1} \and A. Lutovinov\inst{1} \and I. Mereminskiy\inst{1} \and S. Molkov\inst{1} \and M.~Pavlinsky\inst{1} \and B.~D.~Ramsey\inst{5}  \and A. Semena\inst{1} \and N. Semena\inst{1} \and A. Shtykovsky\inst{1} \and R. Sunyaev\inst{1} \and A. Tkachenko\inst{1} \and D. A. Swartz\inst{3} \and G. Uskov\inst{1} \and A. Vikhlinin\inst{1,6} \and  V. Voron\inst{2} \and E. Zakharov\inst{1} \and I. Zaznobin\inst{1}
}

\institute{
    Space Research Institute, 84/32 Profsouznaya str., Moscow 117997, Russian Federation
    \and State Space Corporation Roscosmos, 42 Schepkina str., Moscow 107996, Russia 
    \and Universities Space Research Association, Huntsville, AL 35805, USA
    \and VNIIIEF, Nizhny Novgorod region 607188, Russia
    \and NASA/Marshall Space Flight Center, Huntsville, AL 35812 USA
    \and Harvard-Smithsonian Center for Astrophysics, 60 Garden Street, Cambridge, MA 02138, USA
}

\abstract{We present an updated catalog of sources detected by the \textit{Mikhail Pavlinsky} \art\ telescope aboard the Spektrum-Roentgen-Gamma (SRG) observatory during its all-sky survey. It is based on the data of the first four and the partially completed fifth scans of the sky (\ass). The catalog comprises 1545 sources detected in the 4--12\,keV energy band. The achieved sensitivity ranges between $\sim 4\times 10^{-12}$\,\flux\ near the ecliptic plane and $\sim 7\times 10^{-13}$\,\flux\ near the ecliptic poles, which is a $\sim 30$--50\% improvement over the previous version of the catalog based on the first two all-sky scans (ARTSS12). There are $\sim 130$ objects, excluding the expected contribution of spurious detections, that were not known as X-ray sources before the SRG/\art\ all-sky survey. We provide information, partly based on our ongoing follow-up optical spectroscopy program, on the identification and classification of the majority of the \ass\ sources (1463), of which 173 are tentative at the moment. The majority of the classified objects (964) are extragalactic, a small fraction (30) are located in the Local Group of galaxies, and 469 are Galactic. The dominant classes of objects in the catalog are active galactic nuclei (911) and cataclysmic variables (192). 
}
\keywords{Surveys -- Catalogs -- X-rays: general}

\authorrunning{}

\maketitle

\section{Introduction}
\label{s:intro}

The Spektrum-Roentgen-Gamma (SRG) orbital observatory\footnote{\url{http://srg.cosmos.ru}} \citep{Sunyaev21} was designed to survey the entire sky in X-rays with better sensitivity, angular resolution, and energy coverage compared to its predecessors, such as \uhuru\ (1970--1973), {High Energy Astronomy Observatory 1 (HEAO-1; 1977--1979), and R\"{o}ntgensatellit (ROSAT; 1990--1991). The spacecraft was launched on July 13, 2019, from the Baikonur Cosmodrome to a halo orbit near the L2 point of the Sun--Earth system, from where it started scanning the sky on December 12, 2019. The observatory is equipped with two grazing incidence telescopes: the extended ROentgen Survey with an Imaging Telescope Array (\erosita;  \citealt{Predehl21}) and the \textit{Mikhail Pavlinsky} Astronomical Roentgen Telescope--X-ray Concentrator (\art; \citealt{Pavlinsky21}), which operate in overlapping energy bands of 0.2--8\,keV and 4--30\,keV, respectively. \art\ is a key component of the SRG\ mission because it provides a better sensitivity than \erosita\ at energies higher than 6\,keV, which is particularly important for the systematic search for and exploration of absorbed X-ray sources. Together, the \erosita\ and \art\ all-sky surveys occupy a unique position in the parameter space of X-ray surveys performed thus far (see a recent review by \citealt{2022hxga.book...78B} and in particular Fig.~2 in that paper).

The SRG\ all-sky survey consists of repeat full scans of the sky, each lasting six months. The original plan was to conduct eight such scans. However, on February 26, 2022, the \erosita\ telescope was switched to sleeping mode, and on March 7, 2022, the all-sky survey was interrupted in favor of a deep survey of the Galactic plane region with the \art\ telescope (the Galactic plane and Galactic center regions had also been observed by SRG during the Calibration and Performance Verification phase in 2019, and the resulting catalogs of sources detected by \art\ are presented by \citealt{2023AstL...49..662K} and \citealt{2024MNRAS.529..941S}). This complementary survey was completed in October 2023, after which \art\ resumed the all-sky survey. Hence, by early March 2022, the entire sky had been scanned by SRG\ four times, and about 40\% of the sky had also been covered for a fifth time. 

After completion of the first two all-sky scans (December 2019--December 2020), we compiled a catalog of sources detected by \art\ in the 4--12\,keV energy band (ARTSS12; \citealt{Pavlinsky22})\footnote{Note that due to software issues all ARTSS12 source fluxes and survey flux limits reported in \cite{Pavlinsky22} were underestimated by $\sim 30$\%.}. It comprised 867 X-ray sources, of which nearly 10\% are expected to be spurious due to the design of the catalog.

Here we present an updated version of this catalog, which is based on the entire dataset accumulated by SRG/\art\ during the first $\sim 4.4$ all-sky surveys. The \art\ survey significantly exceeds previous all-sky X-ray surveys carried out in similar energy bands in terms of the combination of angular resolution, sensitivity, and uniformity. Therefore, the presented catalog can be a valuable new source of information for studies of Galactic and extragalactic X-ray source populations.

\section{Data analysis}
\label{s:data}

During the all-sky survey, the optical axes of the \art\ and \erosita\ telescopes are rotating with a period of 4\,hours around the spacecraft $Z$-axis, which is always pointed toward the Sun (this regime of observations is called ``survey mode''). This provides full sky coverage every 6\,months (see \citealt{Sunyaev21} for further details). The catalog of sources presented in this paper is based on the \art\ data accumulated between December 12, 2019, and March 7, 2022, during the first four and the incomplete fifth all-sky surveys, hereafter referred to as \art\ sky surveys 1--5, or \ass\ for short. 

We only used \art\ data that were obtained in survey mode and disregarded data acquired in other (scanning or pointing) observational regimes. We also excised those time intervals when the calibration sources were inserted into the collimators or when high voltage was switched off on the detectors for depolarization \citep{Pavlinsky21}. As the background of the \art\ detector has proved to be exceptionally stable \citep{Pavlinsky21}, virtually no cleaning of the \art\ X-ray data was necessary for periods of high background. Only events detected in one or two upper detector strips and in one or two lower strips\footnote{The coordinate resolution of each \art\ telescope module is provided by two mutually perpendicular sets of 48 strips on the two sides of a CdTe crystal. The strip width corresponds to an angular resolution of 45\arcsec\ \citep{Pavlinsky21}.} were selected. Events detected in a larger number of strips were not used in the analysis because such events are much more likely to be charged particles than photons.

In constructing the \ass\ X-ray map and catalog of sources, we adopted largely the same approach as previously for ARTSS12. Therefore, we refer the reader to \cite{Pavlinsky22} for a detailed description of the different stages of the data analysis. However, we introduced several novel aspects and modifications. In the following subsections, we describe these improvements and briefly outline the components of the analysis that have remained unchanged with respect to the previous version. 

\subsection{Construction of maps}
\label{ss:maps}

As in \cite{Pavlinsky22}, to construct all-sky maps, the \art\ survey data were split into 4,700 overlapping $3.6 \times 3.6$\,deg sky tiles in equatorial coordinates. For each tile, a set of standard maps were prepared, including an exposure map, particle and photon background maps, and sky images. All maps consist of $1024\times 1024$ pixels of 12.66\arcsec. The maps were prepared separately for each of the \art\ surveys (1--5) and then combined. Exposure maps were corrected for vignetting. 

\begin{figure*}
    \centering
    \includegraphics[width=1.5\columnwidth]{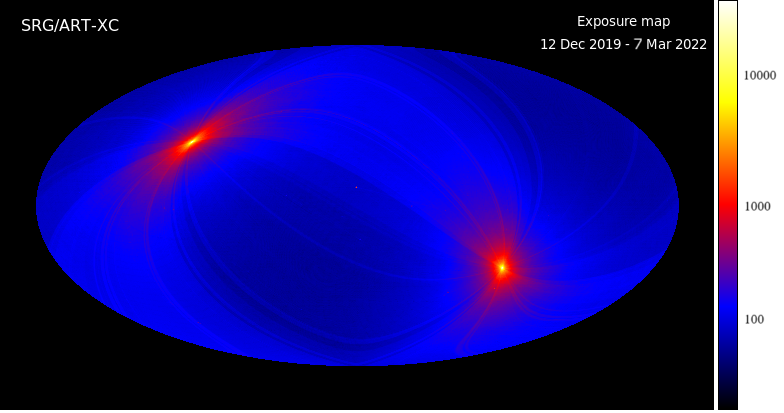}
    \caption{\ass\ exposure map in Galactic coordinates in the 4--12\,keV energy band, with vignetting corrected. The exposure time is given in seconds (see the color scale on the right-hand side).}
    \label{fig:expmap}
\end{figure*}

Figure~\ref{fig:expmap} shows the all-sky vignetting-corrected \ass\ exposure map in the 4--12\,keV band. Due to the strategy of the SRG\ all-sky survey \citep{Sunyaev21}, the accumulated exposure time is lowest near the ecliptic plane, where it varies between $\sim 100$ and $\sim 300$\,s, and highest near the ecliptic poles, where it reaches $\sim 38$\,ks.

\subsection{Background estimation}
\label{ss:bgr}

The SRG/\art\ data are essentially particle background-limited for sources near the detection threshold. As in \cite{Pavlinsky22}, the particle background was estimated using the data in a hard (30--70\,keV) energy band, where the efficiency of the \art\ X-ray optics vanishes. Because the particle background measured in the SRG\ orbit is extremely stable, apart from the gradual multiyear trend associated with the solar cycle activity and rare intense solar flares (see Fig.~\ref{fig:bgr}, based on data from the SRG\ Space Weather Monitor\footnote{\url{https://monitor.srg.cosmos.ru/}}), this method of background estimation is highly robust.

\begin{figure}
    \centering
    \includegraphics[width=1.05\columnwidth]{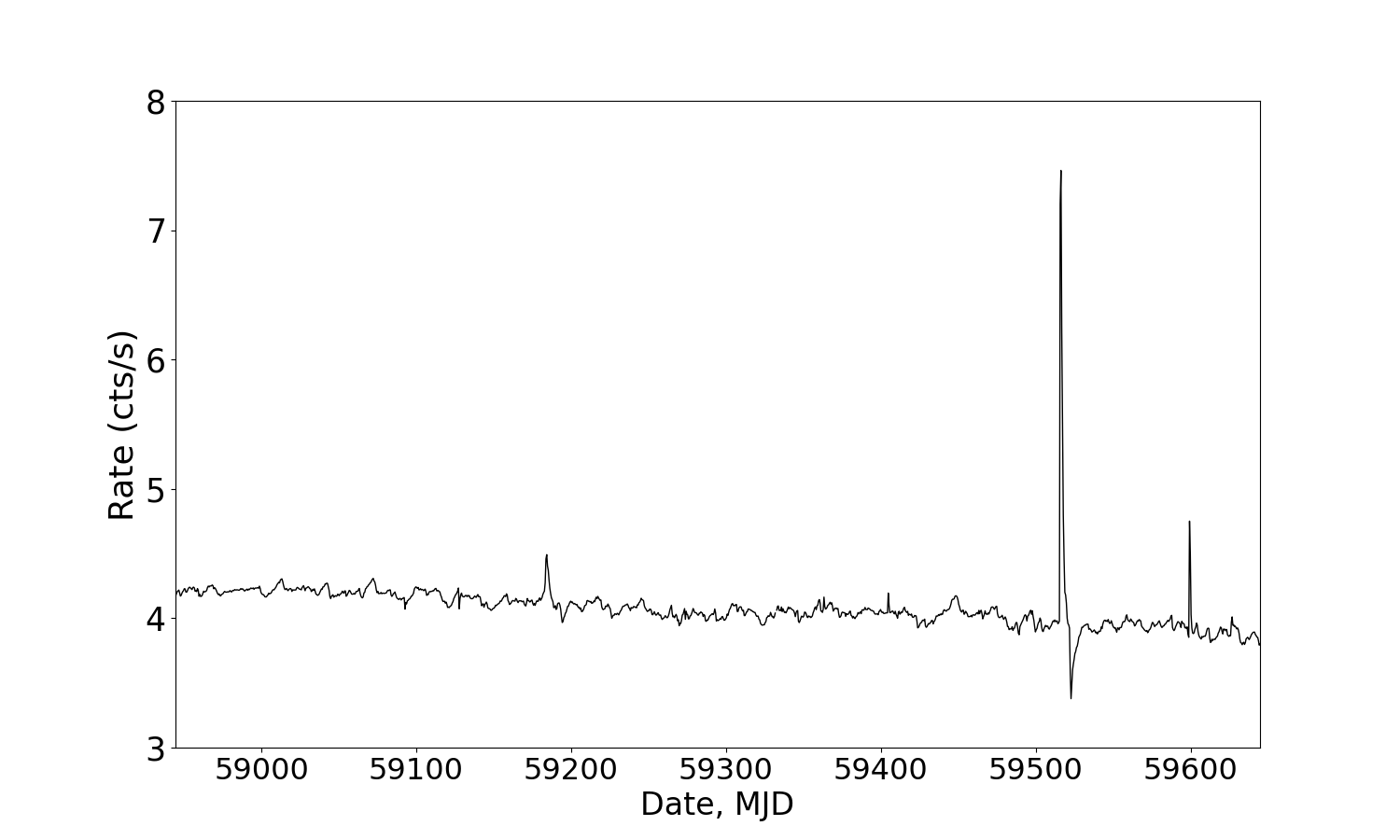}
    \includegraphics[width=1.05\columnwidth]{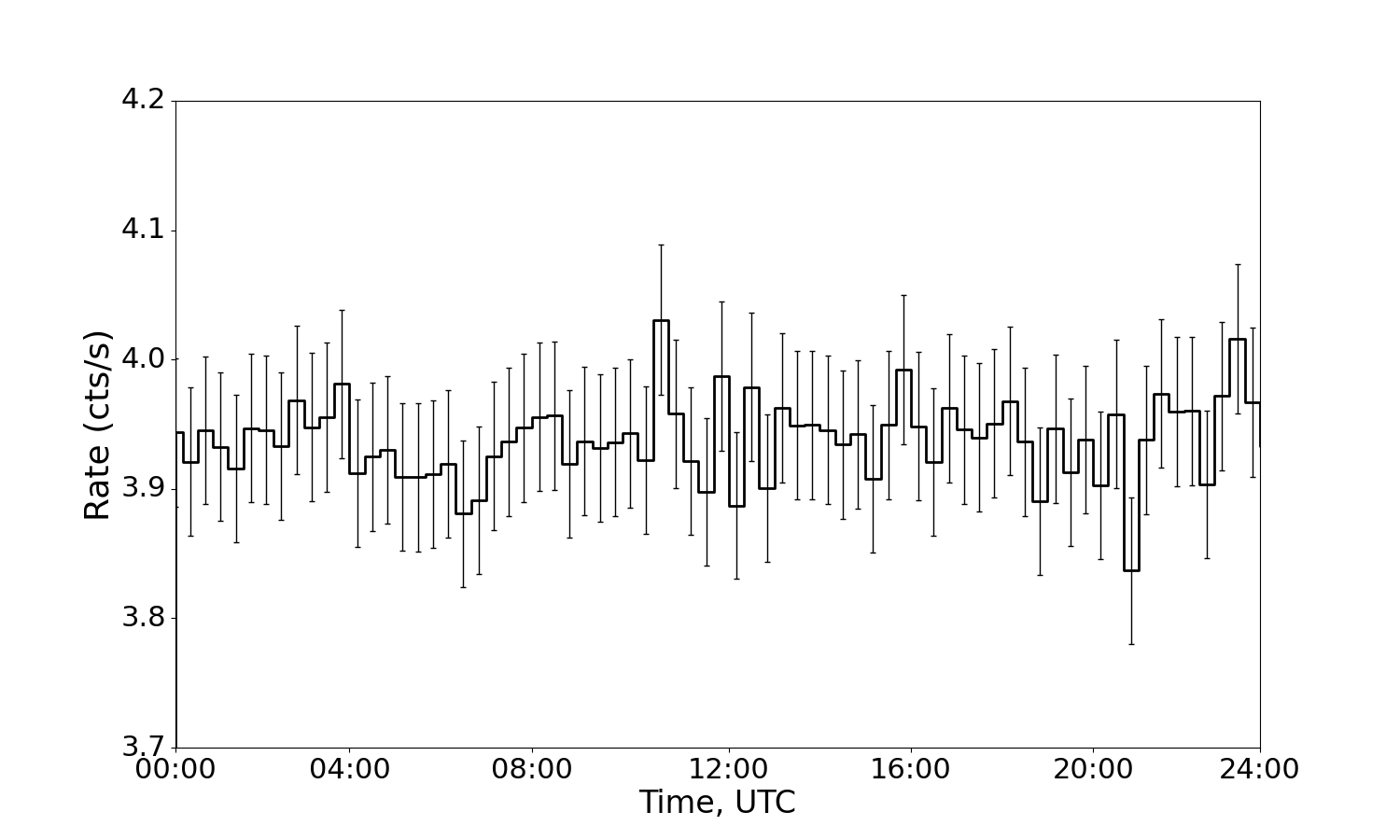}
    \caption{Particle background measured by one of the \art\ detectors in the 60--120\,keV energy band over the course of the SRG\ all-sky survey (top panel) and on an arbitrary day of the survey (bottom panel, with the error bars showing the corresponding Poisson uncertainties). The few sharp variations visible in the long-term light curve are due to solar flares, and such short periods are excised from the data analysis.}
    \label{fig:bgr}
\end{figure}

The cosmic X-ray background provides a negligible contribution ($\sim 3.4$\% in the extragalactic sky) to the total background in the 4--12\,keV energy band and, moreover, the cosmic X-ray background is highly uniform over the sky, except near the Galactic plane and within a few degrees of the Galactic center, where there is strong Galactic ridge X-ray emission (\citealt{2006A&A...452..169R}). However, even in the innermost region of the Galaxy, the contribution of diffuse X-ray emission does not exceed 30\% of the total background and thus has a small effect on the sensitivity to detection of point sources.

The residual background, associated with the uncertainty in particle background estimation and with large-scale X-ray structures in the sky (in particular, Galactic ridge X-ray emission), was assessed from the X-ray images themselves. To this end, the estimated particle background and the point spread function (PSF) models of all bright sources (with X-ray fluxes higher than $\sim 10^{-10}$\,erg\,~s$^{-1}$\,cm$^{-2}$) are subtracted from a given X-ray image, and then all significant details at angular scales smaller than 10\arcmin\ are eliminated from the X-ray image by wavelet decomposition (\texttt{wvdecomp}; \citealt{1998ApJ...502..558V}). 

\subsection{Detection of sources}
\label{ss:det}

We have introduced significant modifications to the source detection procedures compared to the ARTSS12 catalog \citep{Pavlinsky22}. They are explained below. In constructing the \ass\ source catalog, we have focused on the detection of point sources, which constitute the vast majority of sources detected by \art\ during the all-sky survey. We made no attempt to consistently detect extended sources, such as clusters of galaxies or supernova remnants. Nevertheless, many extended sources are still detected with our algorithms designed for the detection of point sources and are thus included in the resulting catalog. However, their estimated X-ray fluxes should be taken with caution.

\subsubsection{Optimal matched filtering}
\label{sss:filter}

In \art\ X-ray images, the noise statistics is close to the Poisson distribution in most cases, and the PSF and vignetting strongly vary across the field of view. We thus adopted the following optimal matched filter, which maximizes the probability of detecting real sources and minimizes the probability of spurious detections of statistical fluctuations:
\begin{equation}
  \Phi(x) = \ln \left[ \frac{f(e) v(x,e) p_s(g)}{b(x,e) p_b(g)} P(x_0 | x) + 1 \right].
  \label{eq:art_phi_x}
\end{equation}
Here, $x_0$ are the photon coordinates, $e$ is the photon energy, $f(e)$ is the expected spectral energy distribution of X-ray sources, $v(x,e)$ is the vignetting function, $b(x,e)$ is the spectral brightness of the background as a function of coordinates and energy, $p_s(g)$ and $p_b(g)$ are the probability distributions of observed event grades (see below) for source photons and background events, respectively, and $P(x_0 | x)$ is the PSF value at $x_0$ under the assumption that the source is located at $x$. 

The optimal matched filter given by Eq.~(\ref{eq:art_phi_x}) thus depends on the energy of every photon, expected source flux, and estimated background at every position in the image. It is applied to \art\ data on an event-by-event basis. We derived this filter as a further development of the Poisson optimal matched filter described in previous works (\citealt{Lynx2019,Ofek18,Pavlinsky22,2024arXiv240402061S}). 

The effective energy band of \art\ in survey observational regime is 4--12\,keV, with sensitivity dropping dramatically below 4\,keV and above 12\,keV. However, because of the finite spectral resolution (FWHM$\sim 1.3$\,keV), a significant fraction of photons with intrinsic energies $\sim 4$\,keV are registered as events with energies between 3 and 4\,keV. For this reason, we used the energy range 3 to 12\,keV for the detection of sources. 

We adopted the same \art\ vignetting function, PSF model, and source spectral shape (namely, a power law with photon index $\Gamma=1.4$) as in our previous work \citep{Pavlinsky22}. In particular, we refer the reader to Sect.~2.3 of \cite{Pavlinsky22} as well as to \cite{2017ExA....44..147K} and \cite{Pavlinsky21} for further details on \art\ PSF calibration and modeling. In particular, Figs.~12-14 in \cite{Pavlinsky21} illustrate the behavior of the PSF at different offset angles and energies. We note that the current PSF model does not take into account the dependence on photon energy. This dependence is only significant in the distant wings of the PSF and is thus more important for offset pointing observations than for the all-sky survey. As regards the background spectrum, we adopted an actually measured spectrum, which was obtained by averaging over the whole set of all-sky survey data.

Each event registered by the \art\ detectors is characterized by a ``grade'' of 0, 1, 2, 3, and so on, which refers to a particular detection pattern in the $3\times3$ pixel area where the event has been registered. The probability of being detected with a given grade is different for photons and charged particles. For example, an event detected in three pixels is most probably a charged particle. We determined the probability distributions $p_s(g)$ and $p_b(g)$ for X-ray sources and background events based on real data of the \art\ all-sky survey.

\begin{figure*}
   \centering
    \includegraphics[width=1.5\columnwidth]{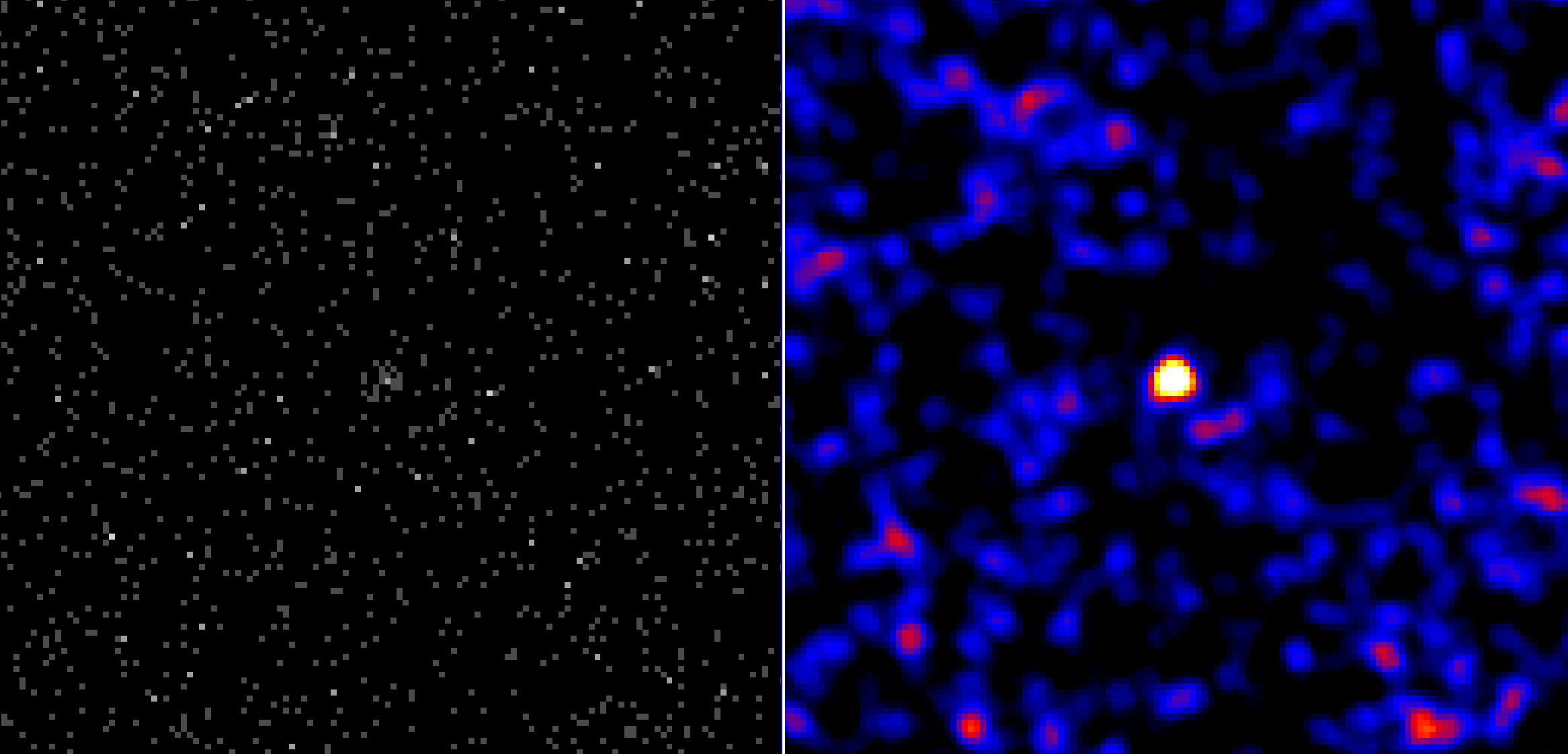}
    \caption{Example of a detection of a faint source using the optimal matched filter given by Eq.~(\ref{eq:art_phi_x}). {\it Left:} Raw photon image in the 4--12\,keV energy band. {\it Right:} Convolution with the filter. The size of the images is approximately $25\arcmin \times 25\arcmin$.}
    \label{fig:conv_img}
\end{figure*}

To determine the flux-to-background ratio involved in the calculation of the optimal matched filter (Eq.~\ref{eq:art_phi_x}), we assumed that point sources have a diameter of 90\arcsec. This region contains approximately 95\% of the photons from a source observed at an offset of less than 10\arcmin, whereas source photons detected from larger offsets are included in the background. The PSF large-angular-scale wings of very bright X-ray sources are also added to the background at this stage. We then adopted the source flux within the 90\arcsec\ diameter aperture that corresponds to a Poisson detection significance of 4.5 for a given background level and substitute this flux for $f(e)$ in Eq.~(\ref{eq:art_phi_x}). The filter thus defined is close to optimal for detecting faint sources (with Poisson detection significance $\gtrsim 4.5)$. For brighter sources, this filter is not optimal but such sources will also be detected with confidence with this filter (see section~A.3.2 in \citealt{Lynx2019}). The filter is only weakly sensitive to the choice of a fiducial source flux.

Sources are detected in images that are obtained by convolution of raw photon images with the optimal filter given by Eq.~(\ref{eq:art_phi_x}) as peaks above some low threshold. Figure~\ref{fig:conv_img} shows an example of a raw image and the resulting ``convolution image''; the latter clearly reveals the presence of an X-ray source. Specifically, we adopted a threshold that yielded a raw catalog of nearly 7000 sources, most of them presumably being spurious detections. At a later stage of the analysis (see below), we adopted a significantly higher threshold, specified in terms of maximum likelihood detection significance and determined by the desired (low) fraction of spurious detections, to construct the final catalog of sources. 

In the case of multiple detections within 52\arcsec\ of each other, that is, within a distance smaller than the \art\ PSF full width at half maximum (FWHM; \citealt{Pavlinsky21}), only the highest peak in the convolution image is taken into account (with the corresponding coordinates and amplitude).

\subsubsection{Maximum likelihood fitting}
\label{sss:ml}

For all the sources detected with optimal matched filtering technique, we performed a maximum likelihood (ML) fit using the following logarithmic likelihood function \citep{1979ApJ...228..939C}:
\begin{equation}
  - 2 \ln L = 2\left(\sum_i \ln m_i - \int m \right),
\end{equation}
where the sum is taken over all the detected events, and the source plus background model is taken as follows:
\begin{equation}
  m (x, e, g) = \sum_j \left[ f(e) v(x,e) p_s(g) P(x_0 | x) +  b(x,e) p_b(g)\right]_j.
\end{equation}
Here, the sum is taken over closely located sources, namely within 5.2\arcmin\ of each other. To fit the model, we used the data within 2.6\arcmin\ of each source; for closely located sources, these data regions were merged. We simultaneously fit the fluxes and coordinates of all sources in the given data region. With this likelihood function, the \art\ PSF and vignetting models as well as information on the event grades are self-consistently taken in account. In reality, the vast majority of sources in the \ass\ catalog were fitted separately. The data regions of two or more sources were merged only in a very small number of cases, because the typical flux threshold for the catalog is much higher than the effective confusion limit.

Sources were sorted by their X-ray fluxes. The brighter ones were fitted first, and their PSF model was then added to the background, including the PSF wings at large angular scales, up to $\approx50$\arcmin. The significance of source detection, that S/N, is calculated from the difference of the logarithmic likelihood function between the best-fit and zero fluxes. The flux and position errors are calculated from the appropriate variation in $ - 2 \ln L$, by varying the parameter of interest with all other parameters frozen at their best fit values.

\subsubsection{Monte Carlo simulations}
\label{sss:mc}

To specify thresholds for source detection in convolution images and validate the detection significance obtained by ML fitting, we carried out Monte Carlo simulations of empty fields. Only the particle background was simulated, because it strongly dominates the \art\ background. Specifically, we computed the expected number of peaks of a given amplitude per square degree in the convolution image of an empty field for different background levels and obtained ML fits for all detected peaks. 

Figure~\ref{fig:sig} shows the expected number of spurious sources per square degree, $\fsp$, as a function of ML detection significance. There is only a weak dependence on the background.

\begin{figure}
   \centering
    \includegraphics[width=\columnwidth]{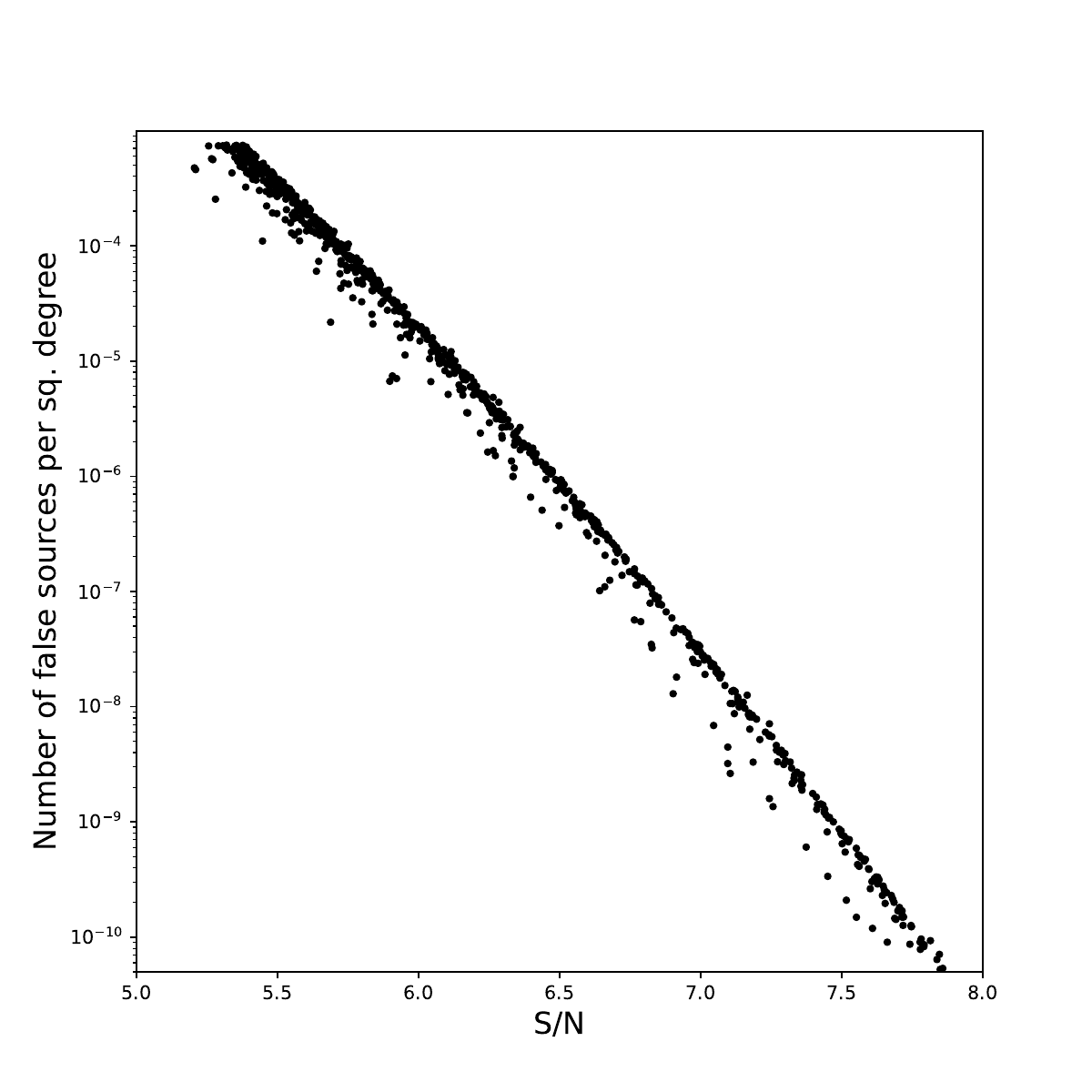}
    \caption{Correlation between the S/N and the expected number of spurious source detections per square degree obtained via simulations of empty fields. The observed scatter reflects a weak dependence on the background.} 
    \label{fig:sig}
\end{figure}

\subsection{Merging sky tiles and calibration of source fluxes}
\label{s:flux}

Using the procedures described above, we obtained catalogs of sources detected in each sky tile. We then merged these individual catalogs into a composite all-sky catalog, taking the overlap of tiles into account. Specifically, sources that are located closer than 3\arcmin\ to the tile edge were rejected and if a source was detected in two or more tiles, the detection at the larger distance from the tile edge was selected. Figure~\ref{fig:img} shows an example of a merged convolution image of a large, crowded region of the sky near the Galactic center.

As was described above, the ML fitting procedure provides the fluxes of detected sources and the corresponding errors. We calibrated the energy flux to count rate conversion factor  for the 4--12\,keV energy band using the available \art\ observations of the Crab nebula (see \citealt{Pavlinsky22} for details). This coefficient in general depends on the adopted procedure of selecting various types of \art\ detector events. For the criteria adopted in this work, a count rate of 1\,cnt~s$^{-1}$ corresponds to a flux of $4.0\times 10^{-11}$\,erg~cm$^{-2}$~s$^{-1}$ in the 4--12\,keV energy band. This takes into account the difference between the slopes of the fiducial source spectrum adopted in this work ($\Gamma=1.4$) and the Crab spectrum ($\Gamma=2.1$). The energy flux to count rate conversion factor  only weakly depends on the spectral shape, namely it varies by less than 4\% for $\Gamma$ between 1.4 and 2.1 and for line-of-sight absorption column densities up to $N_{\rm H}=10^{23}$\,cm$^{-2}$. This is comparable to current uncertainties in the instrument's calibration.

\begin{figure}
   \centering
    \includegraphics[width=\columnwidth]{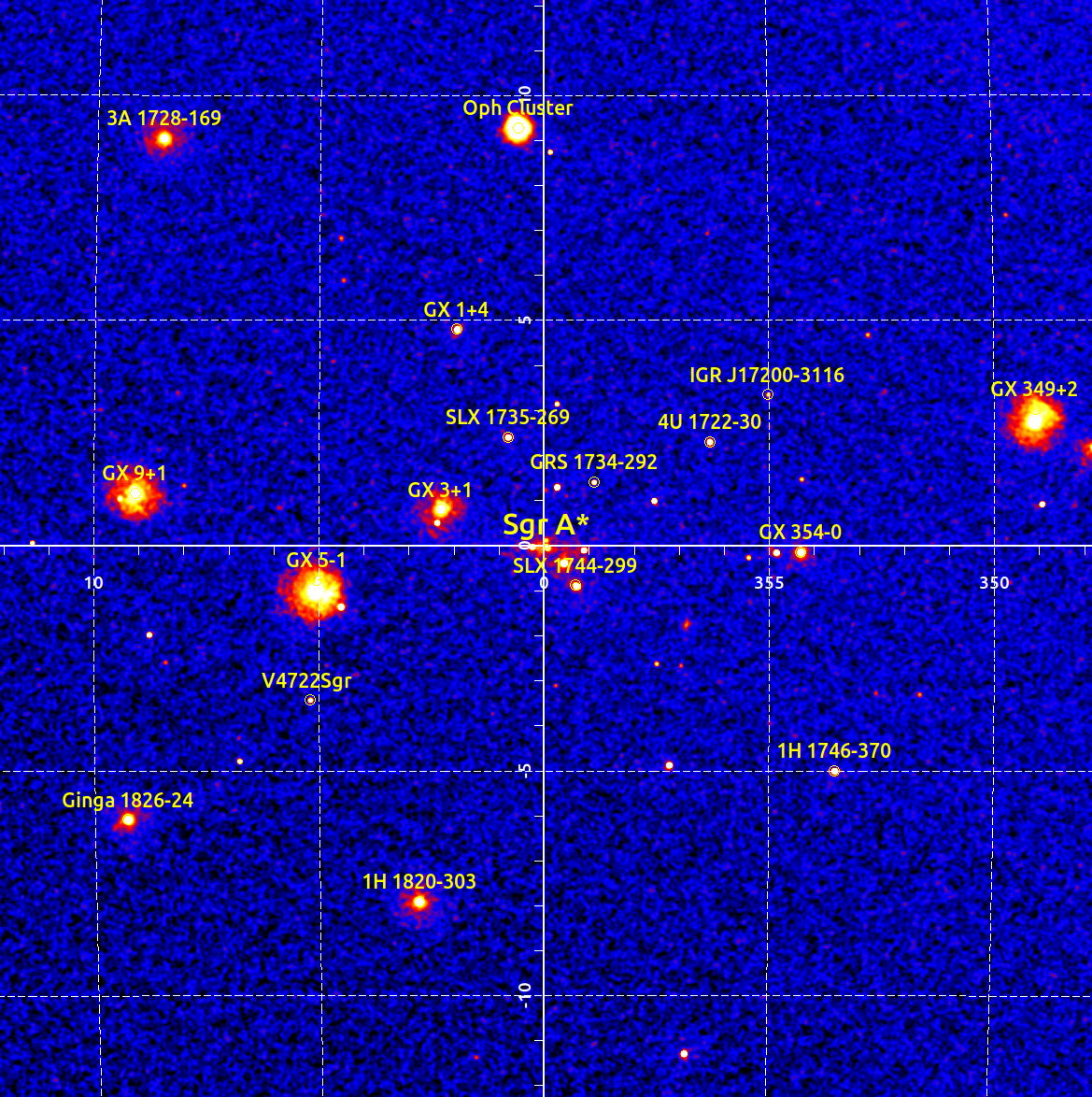}
    \caption{Image of a $25^\circ\times 25^\circ$ region of the sky near the Galactic center in the 4--12\,keV energy band, obtained by convolving the raw photon images with the optimal matched filter, in Galactic coordinates. Some bright sources are labeled. The faintest sources in this region that have been been included in the \ass\ catalog have fluxes $\sim 3\times 10^{-12}$\,\flux.}
    \label{fig:img}
\end{figure}

\section{Construction of the final catalog}
\label{s:final}

To construct the final source catalog, we adopted a threshold of $\fsp=7.475\times 10^{-4}$ spurious sources per square degree, which typically corresponds to an ML detection significance limit of $S/N\approx 5.3$ (see Fig.~\ref{fig:sig}). At this $\fsp$ threshold, the total number of detected sources is 1545, with 2\% of them expected to be spurious. 

\subsection{Identification and classification of \art\ sources}
\label{ss:ident}

For identification of \ass\ sources, we carried out the same kind of analysis as previously for the ARTSS12 catalog \citep{Pavlinsky22}. Namely, we cross-correlated the \ass\ catalog with the SIMBAD Astronomical Database \citep{simbad}, the NASA/IPAC Extragalactic Database\footnote{The NASA/IPAC Extragalactic Database is funded by the National Aeronautics and Space Administration and operated by the California Institute of Technology.} (NED), and the X-ray astronomy database provided by the High Energy Astrophysics Science Archive Research Center (HEASARC). We further searched for possible counterparts in catalogs of optical, infrared (IR), and radio all-sky surveys, in particular \gaia\ Data Release 3 (\gaia\ DR3, \citealt{Gaia2023}), the Sloan Digital Sky Survey (SDSS; \citealt{2020ApJS..249....3A}), the 6dF Galaxy Survey \citep{2009MNRAS.399..683J}, the DECam Plane Survey (DECaPS; \citealt{2018ApJS..234...39S}, the Wide-field Infrared Survey Explorer (WISE) observatory's AllWISE \citep{Cutri2014} and CatWISE2020 \citep{2021ApJS..253....8M} catalogs, the Very Large Array (VLA) Faint Images of the Radio Sky at Twenty-Centimeters Survey (FIRST; \citealt{Helfand2015}), the National Radio Astronomy Observatory (NRAO) VLA Sky Survey (NVSS; \citealt{Condon1998}), the Sydney University Molonglo Sky Survey (SUMSS; \citealt{Mauch2003}), and the Very Large Array Sky Survey (VLASS; \citealt{Gordon21}). 

The search for counterparts was conducted within the 98\%-confidence error radii ($R_{98}$) of \ass\ sources, which does not exceed 30\arcsec\ even for the lowest S/N detections, and within some margin outside these regions. In most cases, the positional precision provided by \art\ enables a straightforward selection of a likely counterpart. Often, previous detections in soft X-rays provide a significantly better localization, which further facilitated this selection. The resulting identifications and classifications as well as redshifts for extragalactic sources were adopted from SIMBAD and/or NED in most cases. 

In cases of dubious identification or classification and for recently or newly discovered X-ray sources, we used additional information from the literature, from our multiwavelength cross-correlation analysis and our optical follow-up campaign (see below). Various photometric (and spectroscopic if available) signatures were taken into account. In particular, the presence of an extended optical object, an IR source with $W1-W2\gtrsim 0.5$ (the color provided by WISE, \citealt{2013ApJ...772...26A}), or a radio source usually indicate an active galactic nucleus (AGN) origin, whereas the presence of a bright star suggests a Galactic nature. All nontrivial cases are discussed on a source by source basis in Appendix~\ref{s:notes}. If the identification and/or classification of a given \ass\ source is not robust, we denote such information as tentative in the catalog. 

\subsection{Follow-up optical program}
\label{ss:follow}

Since the beginning of the SRG\ all-sky survey in December 2019, we have been conducting an extensive program of optical spectroscopy of X-ray sources that have either been discovered by \art\ or were known as X-ray sources from previous missions but have remained unidentified so far. This follow-up campaign is focused on the northern sky (${\rm Dec}>-25^\circ$) and is mostly implemented using two telescopes: the Sayan observatory's 1.6\,m telescope (AZT-33IK; \citealt{2016AstL...42..295B}), operated by the Institute of Solar-Terrestrial Physics of the Siberian branch of the Russian Academy of Sciences, and the Russian-Turkish 1.5\,m telescope (RTT-150), operated jointly by the Kazan Federal University, the Space Research Institute (IKI, Moscow), and the TUBITAK National Observatory (TUG, Turkey). 

This program has already allowed us to identify and classify $\sim 60$ AGN and several cataclysmic variables \citep{2021AstL...47...71Z,2022A&A...661A..39Z,2022AstL...48...87U,2023AstL...49...25U,Uskov24}. In addition, the \ass\ catalog includes a number of Galactic X-ray transients that were discovered during the SRG\ all-sky survey by \erosita\ and/or \art\ and later followed up in the optical (e.g., \citealt{2022A&A...661A..32M,2022A&A...661A..42S}). We took such information into account when assigning identifications and classes to \ass\ sources.

\section{Source catalog}
\label{s:catalog}

\begin{figure*}
    \centering
    \includegraphics[width=1.5\columnwidth]{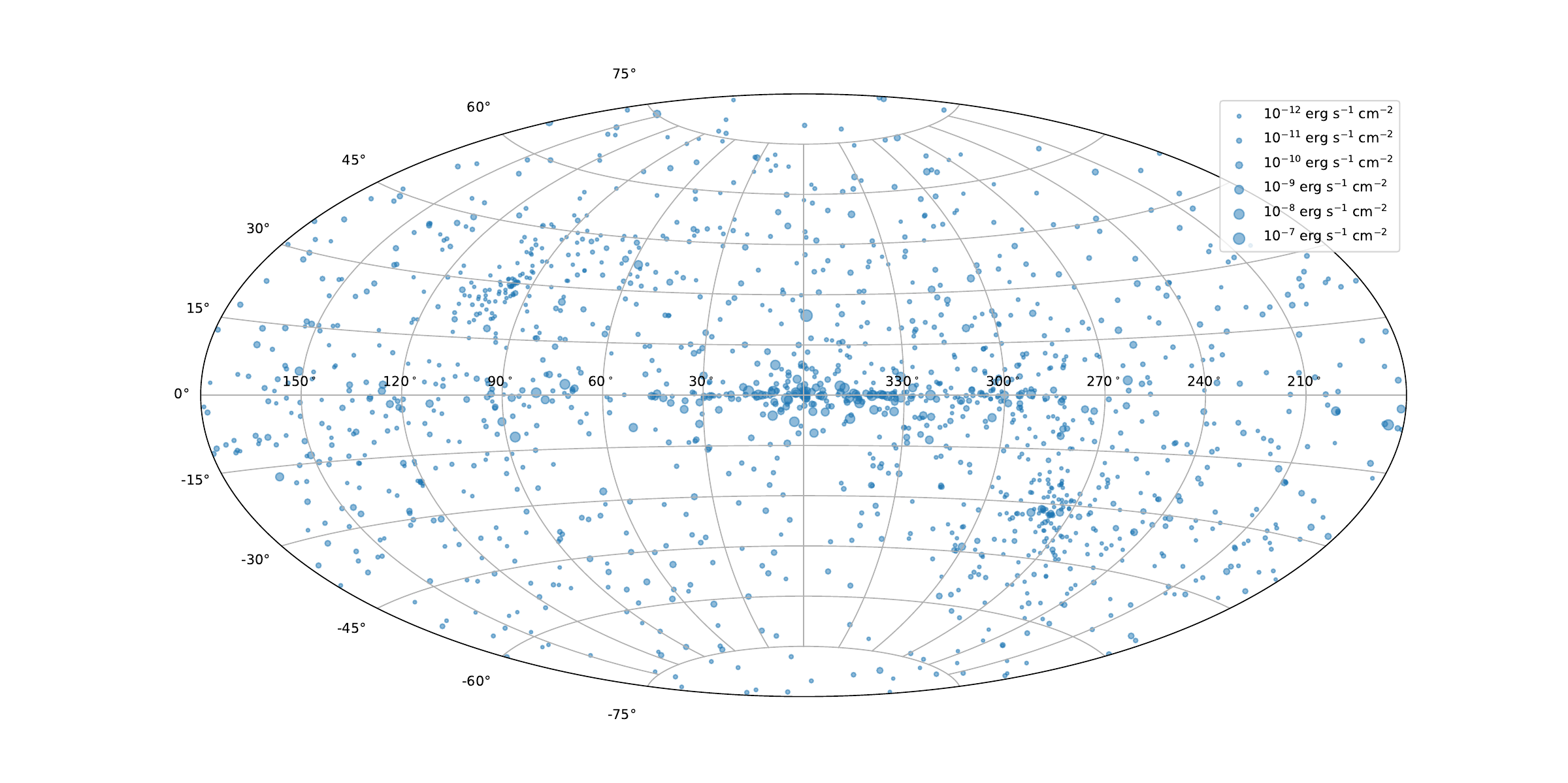}
    \caption{Positions in Galactic coordinates of the X-ray sources detected by \art\ in the 4--12\,keV energy band during \ass. The size of the symbol reflects the X-ray brightness of a source, as indicated in the legend.}
    \label{fig:map}
\end{figure*}

We provide the following information for each source in the \ass\ catalog:

{\it Column (1) ``Id''}: The sequence number of the source in the catalog.

{\it Column (2) ``Name''}: The name of the source in the catalog (prefix ``SRGA'' followed by the source coordinates). 

{\it Columns (3 and 4) ``RA, Dec''}: The equatorial coordinates (J2000).

{\it Column (5) ``R98''}: The statistical position uncertainty at 98\% confidence.  

{\it Columns (6) ``S/N''}: The significance of the detection.

{\it Column (7) ``Flux''}: The flux in the 4--12\,keV energy band and the corresponding 1$\sigma$ uncertainty. 

{\it Column (8) ``Conventional name''}: The conventional name of the source, if available. 

{\it Column (9) ``Redshift''}: The cosmological redshift for extragalactic objects if known. 

{\it Column (10) ``Class''}: The astrophysical class of the object: low- or high-mass X-ray binary (LMXB or HMXB); X-ray binary of uncertain type (X-RAY BINARY); cataclysmic variable or symbiotic binary (CV); ultra-luminous X-ray source (ULX); supernova remnant (SNR); supernova remnant with a central pulsar, when both may contribute to the X-ray emission (SNR/Pulsar); magnetar (MAGNETAR); stellar object, excluding the explicitly mentioned types (STAR); star-forming region (SFR), AGN of the Seyfert or, rarely, LINER type (SEYFERT); unclassified AGN (AGN); beamed AGN (BLAZAR; BL Lac or flat-spectrum radio quasar); cluster of galaxies (CLUSTER); and unclassified source (UNIDENT). A question mark indicates that the quoted classification is tentative (see the comments on these cases in Appendix~\ref{s:notes}).

Building on our previous experience of constructing catalogs of sources detected in all-sky X-ray surveys (e.g., \citealt{2004A&A...418..927R,Krivonos22,Pavlinsky22}), we restricted ourselves to providing a single conventional name for previously known \ass\ sources. To this end, we prefer to indicate the X-ray observatory and/or survey that discovered a given source (e.g., 4U or 2RXS). However, this is often difficult to do, as two or more surveys and/or teams, for example the Burst Alert Telescope (BAT) on board the {\it Neil Gehrels \swift} Observatory and the Imager on-Board the INTEGRAL Satellite on board the INTErnational Gamma-Ray Astrophysics Laboratory (INTEGRAL/IBIS), may have reported the detection of the same source at about the same time. Moreover, many objects were originally known from optical or radio observations rather than from X-ray observations, so in such cases we usually use the optical or radio name (e.g., NGC or 3C). There are further subtleties. In particular, when referring to a source discovered during the ROSAT\ all-sky survey, we prefer to use the name (2RXS) from the (latest) second catalog \citep{Boller16}. However, occasionally we use a shorter (RX or RBS) name if a source is well known under this name from the literature. 

Figure~\ref{fig:map} shows the positions of the \ass\ sources on the sky. The size of the symbols indicates the X-ray brightness of sources.

\subsection{Source counts and survey sensitivity}
\label{s:counts}

\begin{figure}
    \centering
    \includegraphics[width=0.49\textwidth]{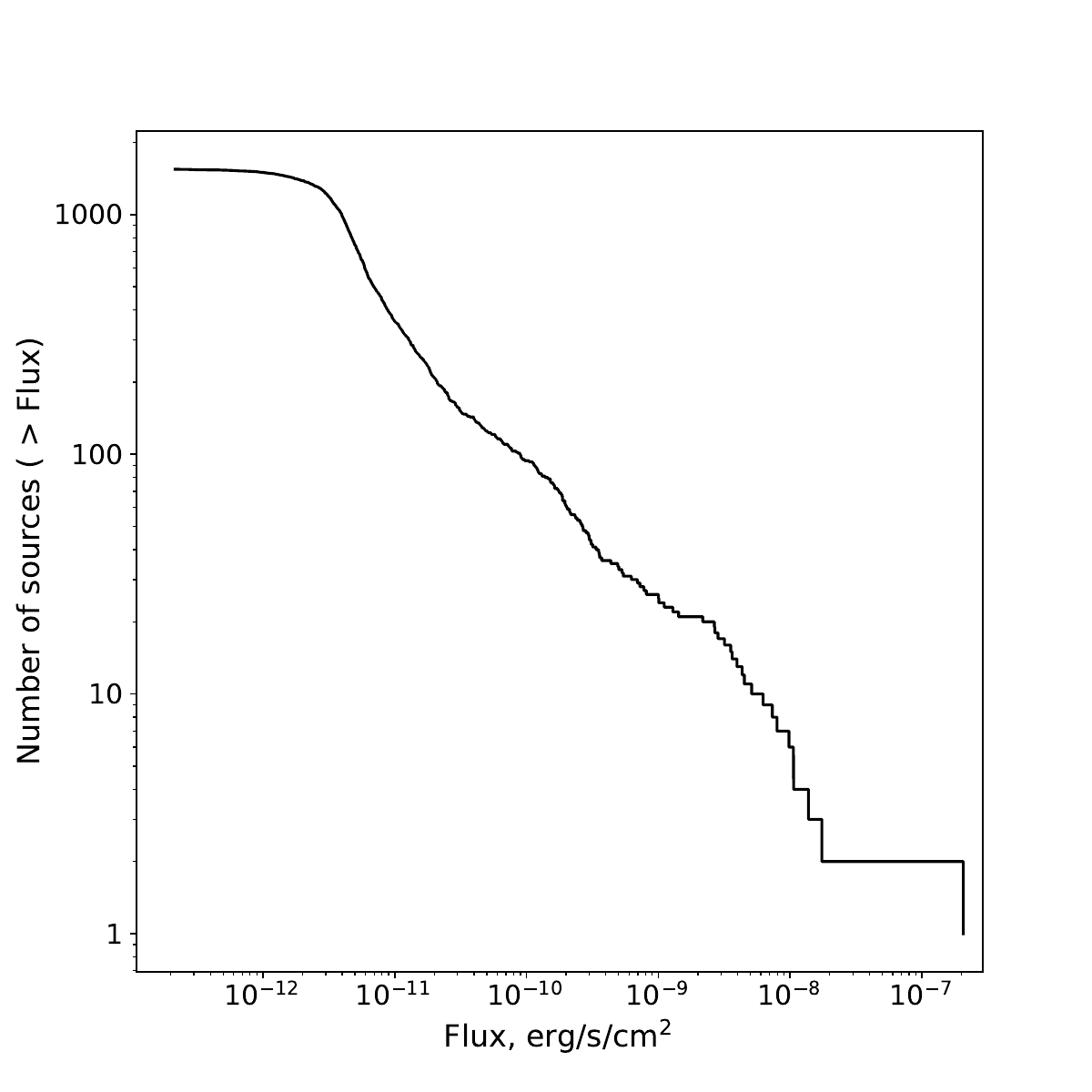}
    \includegraphics[width=0.49\textwidth]{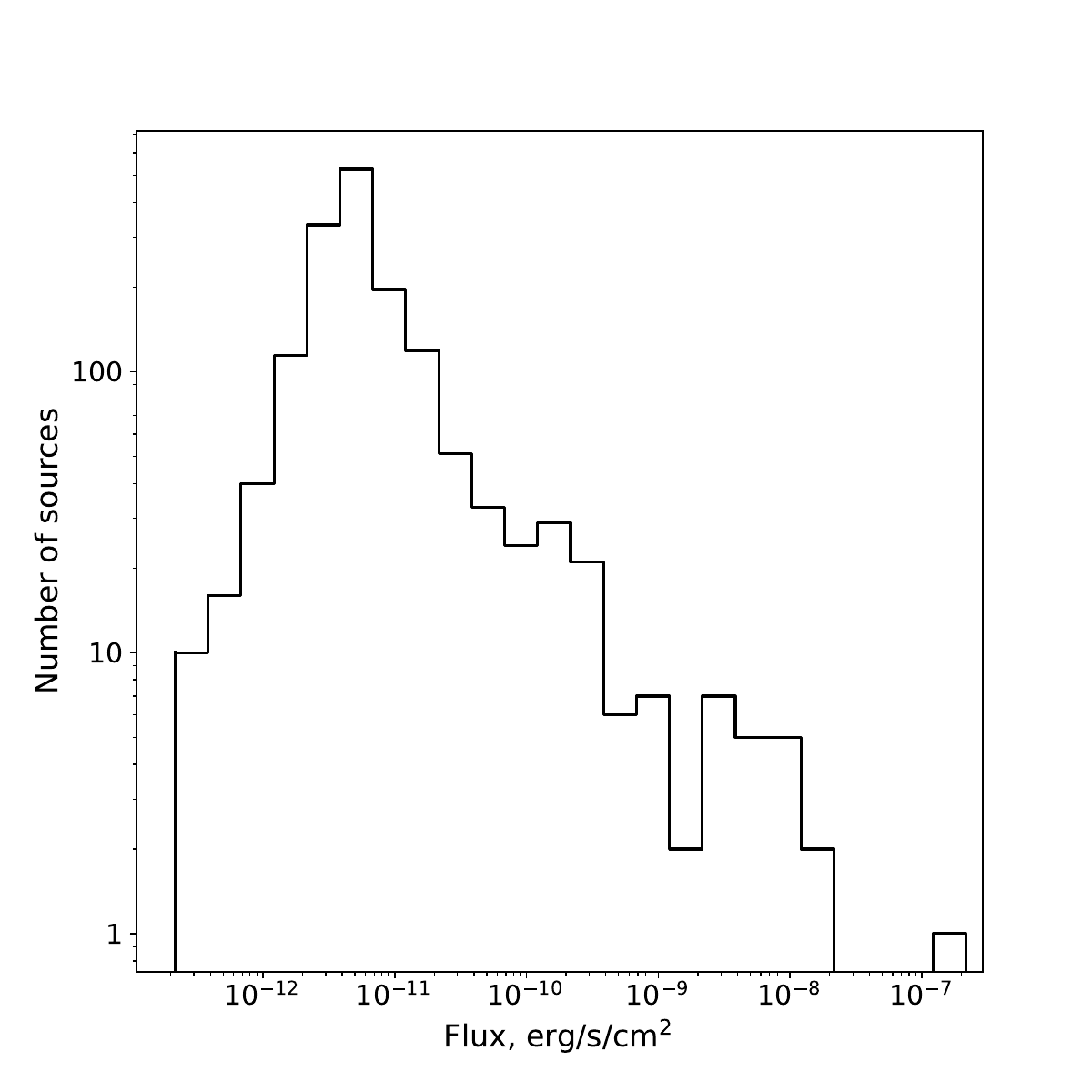}
    \caption{Cumulative (top) and differential (bottom) flux distributions of the \ass\ sources. 
    }
    \label{fig:flux_nsrc}
\end{figure}

Figure~\ref{fig:flux_nsrc} shows the cumulative and differential flux distributions of the \ass\ sources in the 4--12\,keV energy band. The median flux is $4.9 \times 10^{-12}$\,\flux. 

Figure~\ref{fig:b_flux} shows the distribution of the \ass\ sources on the ecliptic latitude--X-ray flux diagram. This plot highlights how the sensitivity of the \art\ all-sky survey monotonically increases from $\sim 4\times 10^{-12}$\,\flux\ near the ecliptic plane (at $|b_{\rm ecl}|<30^\circ$) to $\sim 7\times 10^{-13}$\,\flux\ near the ecliptic poles (at $|b_{\rm ecl}|>82^\circ$). The quoted values are the median fluxes of the sources detected near the detection threshold, namely at $5.3<S/N<6$. An accurate calculation of the sensitivity map of the survey and the completeness of detection of sources as a function of their flux will be presented elsewhere (Burenin et al., in prep.).

\begin{figure}
    \centering
    \includegraphics[width=0.5\textwidth]{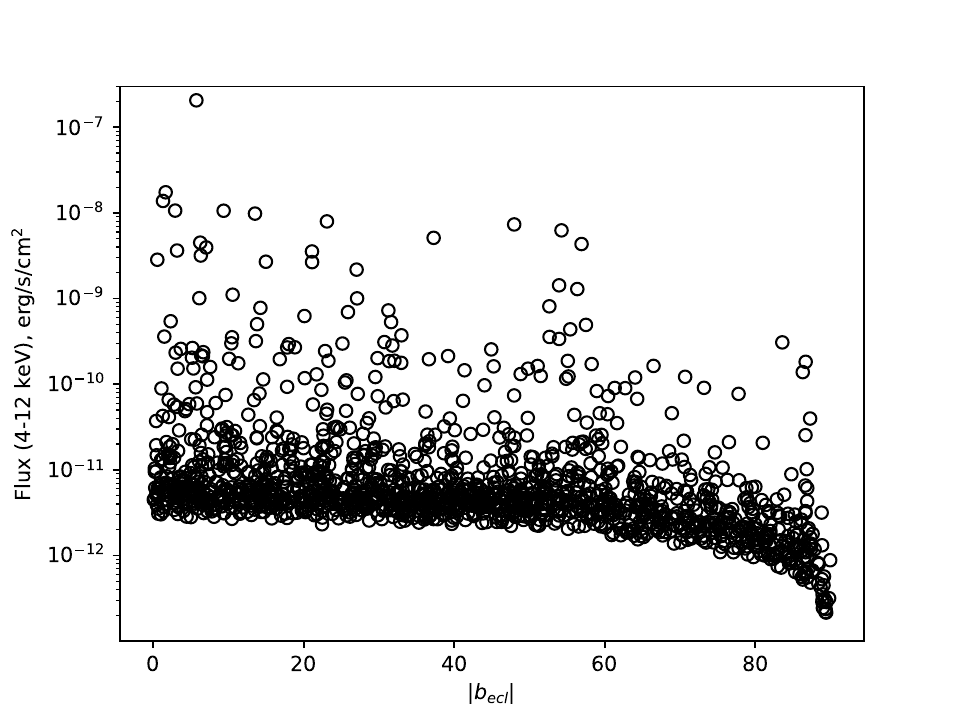}
    \caption{Fluxes of the \ass\ sources as a function of ecliptic latitude. }
    \label{fig:b_flux}
\end{figure}

\subsection{Source classes}
\label{s:class}

\begin{table}
\caption{Statistics of the \ass\ sources by location in the Universe and by type (including tentative classifications).}
\label{tab:source_class}
\centering

\begin{tabular}{cl}
\hline
\hline
Category and type & Count \\
\hline
\hline
Galactic & 469 \\
\hline
LMXB & 97 \\
HMXB & 84 \\
X-ray binary & 2\\
CV & 192 \\
magnetar & 7 \\
star & 66 \\
SNR, SNR/Pulsar & 17 \\
star-forming region & 2 \\
unclassified & 2 \\
\hline
The Local Group & 30 \\
\hline
galaxy & 1 \\
LMXB & 3 \\
HMXB & 19 \\
X-ray binary & 1 \\
ULX & 1 \\
SNR and SNR/Pulsar & 5 \\
\hline
Extragalactic & 964 \\
\hline
galaxy cluster & 48 \\
Seyfert or LINER & 619 \\
blazar & 196 \\
unclassified AGN & 96 \\
galaxy & 1 \\
ULX & 4 \\
\hline
unidentified & 82 \\
\hline
\end{tabular}
\end{table}

Table~\ref{tab:source_class} summarizes the statistics of objects of various classes in the \ass\ catalog. We have managed to identify 1463 out of the 1545 sources. Of these 1463, 174 are tentative identifications, which is to say that for them there is a likely optical/IR counterpart but follow-up observations are needed to ascertain or improve their classification. The majority of the identified objects (66\%) are extragalactic, a small fraction (2\%) are located in the Local Group of galaxies (namely, in the Large Magellanic Cloud, the Small Magellanic Cloud, M31, and M33), and 32\% are of Galactic origin. The largest groups within the Galactic category are CVs (including symbiotic binaries), LMXBs, and HMXBs. The extragalactic objects are dominated by Seyfert galaxies and blazars. 

Extended astrophysical objects such as SNRs and clusters of galaxies are expected to be significantly underrepresented in the \ass\ catalog (for the adopted flux threshold) because our search algorithm is designed for the detection of point X-ray sources. For example, such famous objects as the Coma and Virgo clusters of galaxies are not present in the catalog due to the huge size they subtend on the sky, despite them being bright X-ray sources. 

Eighty-four sources in the catalog, or $\sim 5$\% of all sources, remain unidentified; of these 84 sources, $\sim 30\pm 6$ are expected to be spurious X-ray detections. The majority of the unidentified sources (62) were discovered by \art. The total number of new X-ray sources in the \ass\ catalog (including the identified ones) is 158.  

Figure~\ref{fig:agn_z} shows the redshift distribution of the AGN. The median redshift of the un-beamed (i.e., excluding blazars) AGN is 0.05, and it is 0.25 for the blazars. The redshifts are still missing for 129 AGN and AGN candidates in the \ass\ catalog.

\begin{figure}
    \centering
    \includegraphics[width=0.5\textwidth]{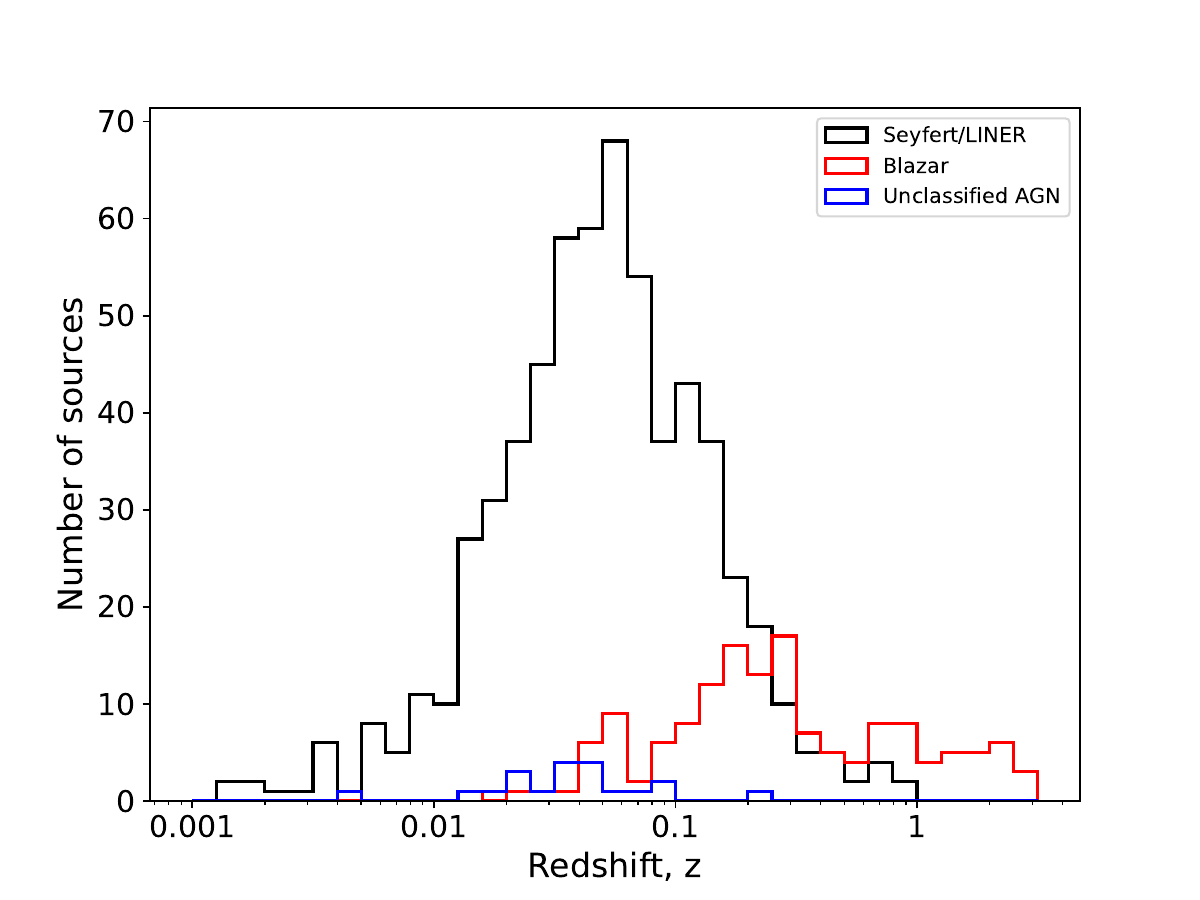}
    \caption{Redshift distribution of the AGN in the \ass\ catalog: Seyfert galaxies (black), blazars (red), and unclassified AGN (blue).}
    \label{fig:agn_z}
\end{figure}

\subsection{Localization accuracy}
\label{s:loc}

\begin{figure}
    \centering
    \includegraphics[width=0.5\textwidth]{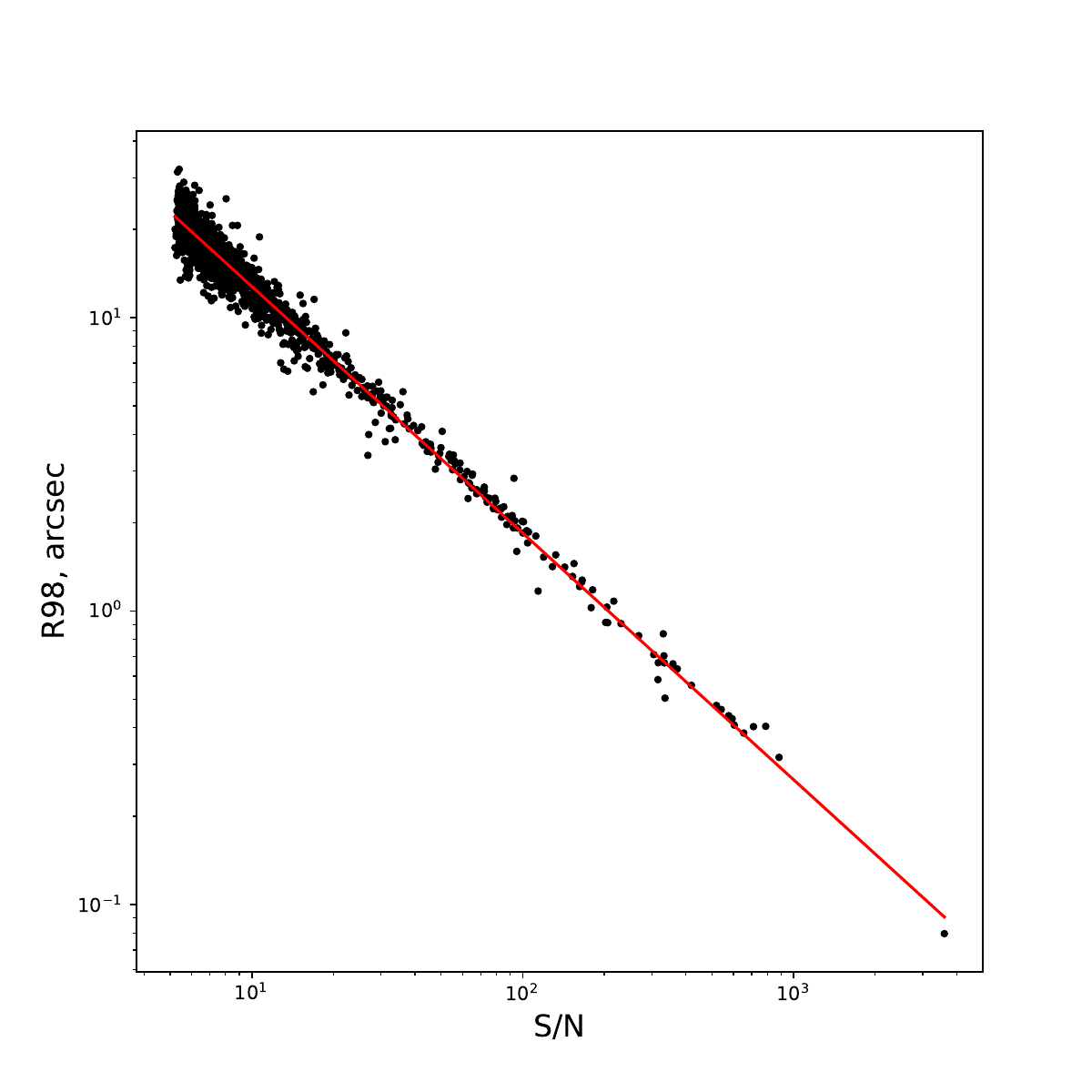}
    \caption{Statistical position uncertainties of the \ass\ sources as a function of their detection significance. The solid line shows the approximate relation between these quantities given by Eq.~(\ref{eq:r98}).}
    \label{fig:r98}
\end{figure}

As already mentioned, we evaluated the statistical uncertainties of source positions using ML fitting. Figure~\ref{fig:r98} shows the resulting $R_{98}$ uncertainties as a function of S/N for the \ass\ sources. There is a clear correlation between these quantities, which can be approximately described as

\begin{equation}
R_{98} \approx 89\arcsec\left(\frac{S}{N}\right)^{-0.84}.
\label{eq:r98}
\end{equation}
The median statistical position uncertainty is $R_{98}=16.1\arcsec$, while for 90\% of the \ass\ sources $R_{98}<22.3\arcsec$.

\begin{figure}
    \centering
    \includegraphics[width=0.5\textwidth]{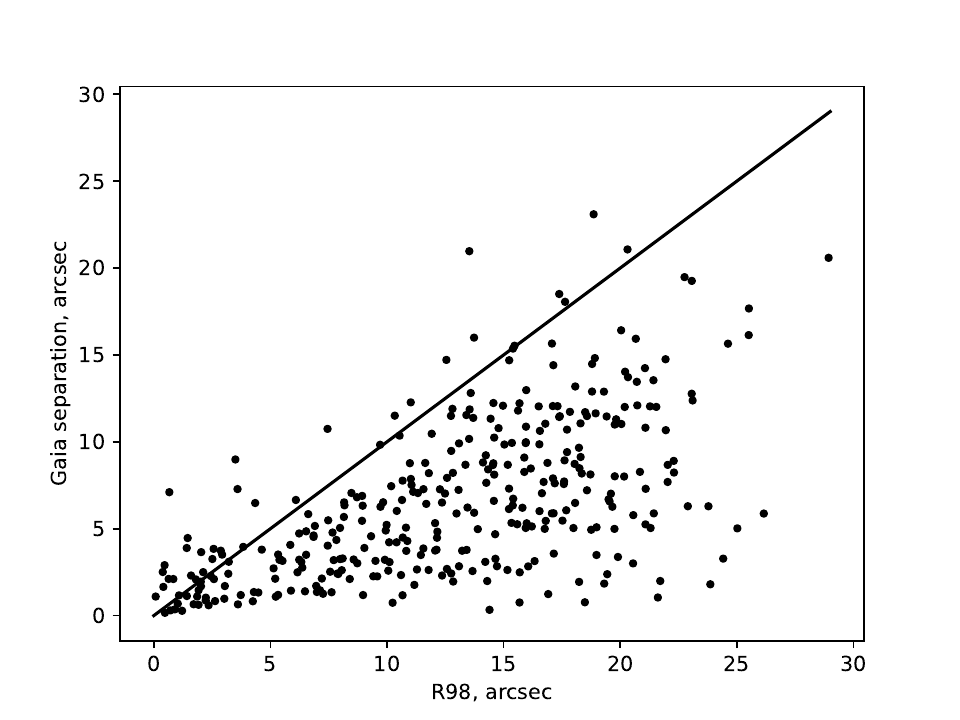}
    \caption{Offsets between the positions of \ass\ sources and their \gaia\ DR3 counterparts versus the 98\%-confidence statistical error radii of the X-ray positions. The diagonal line shows the 1:1 relation.}
    \label{fig:offset}
\end{figure}

There can also be some systematic uncertainties associated with the determination of source positions. To estimate these, we selected those \ass\ sources that can be reliably associated with a point-like Galactic object and have an optical counterpart in the \gaia\ DR3 catalog, and  compared their \art\ and \gaia\ positions. In total, 346 objects were included in this analysis. Figure~\ref{fig:offset} shows the X-ray--optical offsets (${\rm S}$) versus the 98\%-confidence statistical position errors ($R_{98}$) for the test subsample. The fraction of outliers ($S>R_{\rm 98}$), 10\%, exceeds the 2\% expected for the purely statistical uncertainty. Most of these outliers are associated with bright X-ray sources, for which $R_{98}$ is just a few arcseconds. From this comparison, we can estimate a typical systematic positional uncertainty for the \ass\ catalog at $R_{\rm syst}\approx 7\arcsec$, so that the total error radii of \ass\ sources can be crudely estimated as $\sqrt{R_{98}^2+R_{\rm syst}^2}$.

\subsection{Cross-match with external X-ray and gamma-ray source catalogs}
\label{s:cross}

Most of the objects in the \ass\ catalog were known as X-ray sources before. We cross-correlated the catalog with a number of all-sky or nearly all-sky X-ray and gamma-ray surveys: the second ROSAT\ all-sky survey (2RXS; \citealt{Boller16}), the \xmm\ slew survey (XMMSL2; \citealt{Saxton08}, table XMMSLEWCLN in HEASARC), the combined Monitor of All-sky X-ray Image/Gas Slit Camera (MAXI/GSC) 7-year all-sky source catalog (3MAXI; \citealt{Kawamuro18,Hori18}, table MAXIGSC7YR in HEASARC), the \swift/BAT 105-month all-sky survey (hereafter, Swift105mo; \citealt{2018ApJS..235....4O}), the INTEGRAL/IBIS 17-year all-sky survey (hereafter, INT17yr; \citealt{Krivonos22}), and the {\it Fermi} Gamma-ray Space Telescope/Large Area Telescope (\fermi/LAT) 14-year all-sky survey (4FGL-DR4; \citealt{2022ApJS..260...53A,2023arXiv230712546B}). 

We took into account those matches where the angular separation between an \ass\ source and a source from an external catalog was less than the sum of the corresponding 98\% error radii ($R_{98}$). For \ass, we added in quadrature a systematic error of 10\arcsec\ (slightly larger than was estimated in Section~\ref{s:loc}, to be on the safe side) to the statistical errors. The $R_{98}$ values for the external catalogs were computed, assuming a two-dimensional Gaussian probability distribution, from the diverse positional uncertainties provided by these catalogs. Specifically, 2RXS reports 1$\sigma$ positional errors ($\sigma_x$ and $\sigma_y$) on image coordinates, while XMMSL2 and 3MAXI provide 1$\sigma$ error radii ($R_{68}$)\footnote{For a number of 3MAXI sources at low Galactic latitudes without a reported position error in \cite{Hori18}, we estimated $R_{68}$ as 80\arcmin/$S_{\rm det,\,4-10}$, where $S_{\rm det,\,4-10}$ is the \maxi/GSC source detection significance in the 4--10\,keV band.}. For Swift105mo and INT17yr, we estimated $R_{68}$ based on the detection significance of sources, following \cite{2018ApJS..235....4O} and \cite{Krivonos2007}, respectively. The 4FGL-DR4 catalog provides the semimajor and semiminor axes of 95\%-confidence ellipsoidal localization regions; for simplicity, we converted these into 95\%-confidence error radii ($R_{95}$) corresponding to circles of the same area, and also excluded extended 4FGL-DR4 sources. 

We have found 4, 4, and 2 cases where two \art\ sources are blended into one INT17yr, Swift105mo, and 4FGL-DR2 source, respectively. We counted each of these cases as two matches. All of these associations except for two are located in crowded regions of the sky, such as the Galactic plane, Galactic center, and nearby galaxies (Large Magellanic  Cloud, M31, and NGC~4945), where high angular resolution is required to resolve individual X-ray sources. Additionally, there are 98 cases where two or more XMMSL2 sources are found within the match region around an \ass\ source. We regarded each such group as a single match. Most of these cases appear to be caused by imperfect merging of individual \xmm\ slew detections into ``unique sources'' in the XMMSL2 catalog.

Table~\ref{tab:cross_match} provides the results of the cross-matching analysis. The largest overlap of the \ass\ catalog is observed with the 2RXS, XMMSL2, and Swift105mo catalogs. The vast majority of these associations must be real. To estimate the number of spurious  matches, we shifted the positions of the \art\ point sources in random directions by 45\arcmin\ and repeated the cross-matching analysis; we ran many such simulations for each external catalog. The chosen offset is small compared to both the effective angular scale height of the Galactic population of X-ray sources and the effective size of the \art\ deep fields near the ecliptic poles. On the other hand, it is much larger than the positional uncertainties of the vast majority of sources in the external catalogs under consideration, except for a few sources in the 3MAXI and 4FGL-DR4 catalogs. To take these exceptional sources into account, we excluded from the count of spurious matches those few cases where an \ass\ source was linked with the same 3MAXI or 4FGL-DR4 source before and after applying the positional shift (such associations can be real, rather than spurious). In the last column of Table~\ref{tab:cross_match}, we provide the estimated numbers of spurious matches with the external catalogs. 




Of the 1545 \ass\ sources, 194 have not been detected in any of the six all-sky surveys on our list. Taking into account that the \ass\ catalog allows for the presence of $\sim 2$\% of spurious detections (i.e., $\sim 30\pm 6$ sources), we can conclude that $\sim 160$ real sources are unique to the \ass\ catalog among the known all-sky X-ray and gamma-ray catalogs.

No strong trends are apparent in the statistics of cross-identifications with external X-ray source catalogs in terms of celestial position or source class. Specifically, we compared the relative fraction of cross-matches of \ass\ sources with the 2RXS, XMMSL2, and Swift105mo catalogs (i) near and outside the Galactic plane ($|b|<10^\circ$ versus $|b|>10^\circ$) and (ii) for Galactic versus extragalactic sources (according to Table~\ref{tab:source_class}), and the resulting numbers differ just by several per cent for each of the mentioned external catalogs. In particular, for 2RXS, the largest difference in the relative (with respect to \ass) fraction of cross-matches is found between Galactic and extragalactic sources: 57.4\% versus 69.0\%, whereas for Swift105mo the largest difference is observed between the $|b|<10^\circ$ and $|b|>10^\circ$ samples, namely 61.7\% versus 49.8\% (for XMMSL2 the variations are yet smaller). It is difficult to interpret this information, because all these surveys differ significantly in their spectral response and coverage of the sky (expect for 2RXS, whose exposure map is skewed toward the Ecliptic poles as for SRG/\art) and, moreover, XMMSL2 has not covered the whole sky. However, intrinsic variability, spectral hardness, and line-of-sight absorption (both intrinsic and Galactic) certainly play key roles in this statistics. 

We note that all of the surveys considered here, except for 2RXS, were conducted over periods of at least 8 years but some of them (in particular, XMMSL2) are characterized by a low duty cycle. For comparison, the \ass\ catalog is based on a relatively short period of 2.3 years of nearly continuous scanning of the sky, characterized by a duty cycle of more than 97\%.

\begin{table*}
\caption{Cross-match of the \ass\ sources with selected X-ray and gamma-ray source catalogs.}
\label{tab:cross_match}
\centering

\begin{tabular}{ccrrr}
\hline
X-ray survey & Energy band & Reference & Cross-  & Spurious \\
             &             &           & matches & matches$^{a)}$ \\
\hline
ROSAT\ (2RXS) 1 year & 0.1--2.4\,keV & \citet{Boller16} & 960 & 4.5 (2--7)  \\
\xmm\ slew & 0.2--12\,keV & \citet{Saxton08} & 874  &  1.8 ($<4$) \\
MAXI/GSC 7 years  & 4--10\,keV & \citet{Kawamuro18,Hori18} & 286  & 5.6 (3--9) \\
\swift/BAT 105 months & 14--195\,keV & \citet{2018ApJS..235....4O} & 785  &  5.8 (3--8) \\
INTEGRAL\ 17 years & 17--60\,keV & \citet{Krivonos22} & 504  &  4.5 (2--7)  \\
\fermi/LAT 14 years & 50 MeV--1 TeV & \citet{2022ApJS..260...53A} & 222   & 23.8 (18--30) \\
\hline
\end{tabular}
\begin{flushleft}    
$^{a)}$ Expected value of spurious matches and the corresponding 90\% confidence interval.
\end{flushleft}
\end{table*}




\subsection{Ecliptic poles}
\label{s:ecl}

Longer exposures were carried out for the regions around the ecliptic poles compared to other fields during the \art\ all-sky survey (see Fig.~\ref{fig:expmap}), and it is interesting to examine the composition of the X-ray source samples detected in these ``deep surveys.'' Specifically, we focus our attention on the regions $b_{\rm ecl}>82^\circ$ and $b_{\rm ecl}<-82^\circ$, with an area of $\approx 200$\,square degrees each, around the north and south ecliptic poles (NEP and SEP), respectively. For the adopted detection significance threshold, the expected number of spurious sources in the combined NEP--SEP sample is $\sim 0.3$ (i.e., this sample is expected to be highly pure).

There are 49 \ass\ sources, with fluxes down to $\sim 2\times 10^{-13}$\,\flux, in the NEP field. The majority of them (44) were known as X-ray sources before and 5 sources have been discovered by \art. These five sources proved to be AGN, based on our follow-up optical spectroscopy program (in addition, we identified via optical spectroscopy 3 AGN among the previously known X-ray sources). Most of the NEP sample are extragalactic sources: 3 galaxy clusters and 41 AGN (including candidates). There are also 5 Galactic objects: 3 CVs and 2 coronally active stars. 

The SEP region contains 60 \ass\ sources, with fluxes down to $\sim 3\times 10^{-13}$\,\flux, including nine X-ray sources discovered by \art. One-third of the SEP sample (20 objects) are X-ray binaries and SNRs residing in the Large Magellanic Cloud. Most of the remaining sources are extragalactic: one galaxy cluster and 33 AGN (including candidates). There are also five Galactic objects (three CVs and CV candidates, one HMXB, and one star), and one unclassified object. As expected, the AGN detected in the NEP and SEP fields are on average more distant than those found elsewhere, with the most distant un-beamed AGN located at $z\sim 0.9$.  

\section{Discussion and summary}

We have updated the catalog of sources detected by the \art\ telescope during the SRG\ all-sky survey by adding data from the third, fourth, and the $\sim 40$\% complete fifth half-year scans of the sky to the data of the first two scans on which the previous version of the catalog (ARTSS12, \citealt{Pavlinsky22}) was based. The new catalog comprises 1545 sources detected in the 4--12\,keV energy band. The achieved sensitivity ranges between $\sim 4\times 10^{-12}$\,\flux\ near the ecliptic plane and $\sim 7\times 10^{-13}$\,\flux\ near the ecliptic poles. The new \art\ catalog is $\sim 1.3$--1.5 times deeper than the previous one. The sensitivity as a function of celestial coordinates will be evaluated more accurately elsewhere (Burenin et al., in prep.), based on extensive numerical simulations of the survey. 

Just a few all-sky or nearly all-sky surveys have been conducted in similar medium X-ray bands (i.e., at energies $\sim 2$--20\,keV). In particular, the HEAO-1\ experiment A2 performed a survey of the extragalactic sky ($|b|>20$\,deg) in the 2--10\,keV energy band \citep{1982ApJ...253..485P}, the {\it Rossi} X-ray Timing Explorer (RXTE) Slew Survey (XSS) covered the $|b|>10^\circ$ sky in the 3--20\,keV band, and the MAXI/GSC all-sky survey (3MAXI) covered the entire sky in the 4--10\,keV band \citep{Kawamuro18,Hori18}. All these surveys had much lower angular resolutions ($\sim 1^\circ$) compared to \ass, which led to source confusion in crowded regions of the sky and complicated the identification of X-ray sources. The median sensitivities of the HEAO-1\ A2 survey, XSS, and 3MAXI, converted to the 4--12\,keV energy band, are $\sim 1.8\times 10^{-11}$, $\sim 1.1\times 10^{-11}$, and $\sim 8\times 10^{-12}$\,\flux\ (for spectra not very different from those of the Crab), which is significantly worse than the sensitivity achieved during \ass\ in its shallowest part near the ecliptic plane. 

The \xmm\ slew survey (XMMSL2; \citealt{Saxton08}) covered $\sim 84$\% of the sky\footnote{\url{https://www.cosmos.esa.int/web/xmm-newton/xmmsl2-ug}}. The median flux of the sources detected in its hard band of 2--12\,keV is $9.3\times 10^{-12}$\,\flux, which corresponds to $\sim 5.7\times 10^{-12}$\,\flux\ in the 4--12\,keV band for Crab-like spectra. This should be compared to the median flux of $\sim 4.9\times 10^{-12}$\,\flux\ of the \ass\ sources. Therefore, XMMSL2 (hard band) and \ass\ are comparable in sensitivity, but the latter provides a more regular coverage of the sky and is done in a slightly harder energy range.

As a result of the increased depth compared to the first year of the \art\ all-sky survey \citep{Pavlinsky22}, the dominance of AGN among the classified sources has strengthened, with $\sim 60$\% of the \ass\ catalog sources being AGN or AGN candidates. The catalog includes 158 objects that were not known as X-ray sources before. Given that $\sim 30$ of them are expected to be spurious detections (for the adopted detection significance threshold), there are $\sim 130$ truly new X-ray sources, which corresponds to $\sim 8$\% of the entire catalog. This fraction has increased with respect to ARTSS12, as anticipated.

The \ass\ catalog has a significant added value in terms of the identification and classification of sources. Nearly 83\% of the sources are already reliably classified and another $\sim 11$\% have tentative classifications. We are pursuing the goal of achieving a nearly 100\% identification completeness. To this end, we have been carrying out a follow-up optical spectroscopy program, which has already allowed us to identify and classify $\sim 60$ AGN and $\sim 10$ Galactic objects (mostly CVs). So far, this campaign has been focused on the northern sky (${\rm Dec}>-25^\circ$), and we urge its extension to the southern hemisphere. Our ultimate objective is to provide statistically complete samples of different classes of objects, in particular AGN and CVs, selected in the 4--12\,keV energy band, which is largely unique to the \art\ survey. The current samples of AGN and CVs in the \ass\ catalog comprise $\sim 900$ and $\sim 200$ objects, respectively. We are now developing an \ass\ web database\footnote{to be located at \url{https://www.srg.cosmos.ru}} that will provide additional information on the counterparts of the \ass\ sources.

The SRG/\art\ all-sky survey has recently been resumed after a 1.5-year pause. Four new scans of the sky are to be conducted by the end of 2025. The next official release of the \art\ catalog should be based on all eight scans of the sky, and according to our preliminary estimates the final catalog will contain more than 2500 sources.

\begin{acknowledgements}
The \textit{Mikhail Pavlinsky} \art\ telescope is the hard X-ray instrument on board the SRG\ Observatory, a flagship astrophysical project of the Russian Federal Space Program realized by the Russian Space Agency in the interests of the Russian Academy of Sciences. The \art\ team thanks the Russian Space Agency, Russian Academy of Sciences, and State Corporation Rosatom for the support of the SRG\ project and \art\ telescope. We thank Lavochkin Association (NPOL) with partners for the creation and operation of the SRG\ spacecraft (Navigator). We thank Acrorad Co., Ltd. (Japan), which manufactured the CdTe dies, and Integrated Detector Electronics AS -- IDEAS (Norway), which manufactured the ASICs for the X-ray detectors. SS, RB, EF, RK, IM, GU, EZ, and IZ acknowledge the support of this research by the Russian Science Foundation (grant 19-12-00396). We are grateful to the referee for the helpful comments and suggestions.
\end{acknowledgements}

\bibliographystyle{aa} 
\bibliography{artss15}

\begin{appendix} 

\section{Notes on individual sources}
\label{s:notes}

Here\footnote{\url{https://zenodo.org/doi/10.5281/zenodo.11191704}}, we provide additional comments on the identification and classification of a number of sources from the \ass\ catalog, namely those with dubious identification or classification as well as recently or newly discovered X-ray sources. In addition to the databases and catalogs already mentioned in Section~\ref{ss:ident}, we acquired this information from the literature and from our ongoing follow-up optical spectroscopy campaign. 

For a few of the objects discussed below, we obtained a tentative Seyfert classification (class ``Seyfert?'' in the catalog) by inspecting their optical spectra available from the SDSS or 6dF surveys. To this end, we used the standard criteria based on the widths and intensities of emission lines \citep{Osterbrock81,Baldwin81} following our previous work on the classification of AGN from the SRG/\art\ all-sky survey (e.g., \citealt{2023AstL...49...25U,Uskov24}).


\input{notes/artxc_paper2_notes_20240322-115342}

\input{longtable_arxiv}

\end{appendix}

\end{document}

%% file: notes/artxc_paper2_notes_20240322-115342.tex
\subsection*{ SRGA J000048.1-070914 }
A Seyfert 1.9 at $z=0.0375$ \citep{2022ApJS..261....2K}.
\subsection*{ SRGA J000132.9+240237 }
Associated with 2MASX\,J00013232+2402304, a Seyfert 1.9 at $z=0.1048$ \citep{Uskov24}.
\subsection*{ SRGA J001439.6+183500 }
A Seyfert 2 at $z=0.0180$ \citep{2023AstL...49...25U}.
\subsection*{ SRGA J002203.6+254017 }
A Seyfert 1.2 at $z=0.1292$ \citep{2017MNRAS.468..378S}.
\subsection*{ SRGA J002241.4+804348 }
A Seyfert 1 at $z=0.1147$ \citep{2023AstL...49...25U}.
\subsection*{ SRGA J003535.1+462345 }
Associated with the dwarf nova GALEX\,J003535.7+462353 \citep{2003A&A...404..301R,2011IBVS.5982....1W}.
\subsection*{ SRGA J003947.5-754949 }
Likely associated with the galaxy LEDA\,242675, $W1-W2\sim 1.1$.
\subsection*{ SRGA J004142.9+413410 }
Likely an LMXB in the globular cluster Bol 45 in M31 \citep{2011A&A...534A..55S}.
\subsection*{ SRGA J004223.1+290401 }
Associated with the known Seyfert galaxy 2MASX\,J00422385+2903588 at $z=0.0711$ \citep{2018MNRAS.474.1873W}, previously undetected in X-rays.
\subsection*{ SRGA J004238.9+411603 }
The central region of M31, unresolved into individual X-ray sources.
\subsection*{ SRGA J004313.9+410723 }
In M31, might be confused with other sources.
\subsection*{ SRGA J004505.9+620746 }
Likely an SNR (Mereminskiy et al., in prep.).
\subsection*{ SRGA J004545.6+413956 }
An X-ray binary in the globular cluster Bo 375 in M31, likely with a neutron star \citep{2008ApJ...689.1215B,2016MNRAS.458.3633M}.
\subsection*{ SRGA J004732.9-251713 }
A starburst galaxy with many point X-ray sources and hot interstellar gas.
\subsection*{ SRGA J004852.9-734945 }
In the SMC.
\subsection*{ SRGA J005456.2-722644 }
In the SMC.
\subsection*{ SRGA J005642.6+604301 }
The prototype gamma Cas star.
\subsection*{ SRGA J010336.0-720135 }
In the SMC.
\subsection*{ SRGA J010430.3-723133 }
In the SMC.
\subsection*{ SRGA J010743.0+574423 }
A Seyfert 1.9 at $z=0.0699$ \citep{2023AstL...49...25U}.
\subsection*{ SRGA J011506.0+882903 }
Likely associated with the bright star Gaia DR3  
576260267526450048, at a distance $\sim 400$~pc, a spectroscopic binary in a young stellar association \citep{2020A&A...637A..43K}. Hence, likely a coronally active star.
\subsection*{ SRGA J011630.5-123557 }
A Seyfert 1.9 at $z=0.1424$ \citep{2022ApJS..261....2K}.
\subsection*{ SRGA J011704.8-732638 }
In the SMC.
\subsection*{ SRGA J011905.4+160504 }
Associated with 2MASX\,J01190583+1604550, a Seyfert 2 galaxy at $z=0.0700$ (Uskov et al., in prep.).
\subsection*{ SRGA J012823.7+162743 }
A Seyfert 2 at $z=0.0383$.
\subsection*{ SRGA J013350.9+303932 }
In M33.
\subsection*{ SRGA J014458.7-023211 }
Likely associated with the galaxy 2MASX\,J01445853-0231589, $W1-W2\sim 0.7$.
\subsection*{ SRGA J015006.8-322444 }
Possibly associated with the galaxy 
LEDA\,131950 at $z=0.0384$, $W1-W2\sim 0.3$.
\subsection*{ SRGA J015440.4-270705 }
A Seyfert 1 at $z=0.1510$ \citep{2022ApJS..261....2K}.
\subsection*{ SRGA J015524.4+022825 }
A Seyfert 1 at $z=0.0847$ \citep{2022ApJS..261....2K}.
\subsection*{ SRGA J015641.5-835833 }
Likely associated with the star Gaia DR3 4617143036371460864 at a distance $\sim 500$\,pc.
\subsection*{ SRGA J015657.4-530202 }
The spectroscopic redshift is $z=0.3043$ \citep{2021A&A...650A.106G}.
\subsection*{ SRGA J020012.7+065507 }
Associated with the bright, long-period variable star DE Psc at a distance $\sim 700$\,pc. Hence, possibly a symbiotic star.
\subsection*{ SRGA J020639.4-714824 }
A flat-spectrum radio source \citep{2007ApJS..171...61H}.
\subsection*{ SRGA J020749.4+445032 }
A Seyfert 1.9 at $z=0.0217$ \citep{2022ApJS..261....2K}.
\subsection*{ SRGA J021643.4+255250 }
Associated with the galaxy LEDA\,1753646, a Seyfert 2 at $z=0.0663$ (Uskov et al., in prep.).
\subsection*{ SRGA J022235.4+250818 }
A Seyfert 1 at $z=0.0616$ \citep{2022ApJS..261....2K}.
\subsection*{ SRGA J022625.5+592746 }
A radio galaxy with unknown AGN optical type and redshift  \citep{2021MNRAS.500.3111B}.
\subsection*{ SRGA J022746.2-693129 }
A giant star at a distance $\sim 240$\,pc. The exact nature of the X-ray emission is unknown.
\subsection*{ SRGA J023800.1+193818 }
Associated with 2MASS\,J02375999+1938118, a Seyfert 1 galaxy at $z=0.0335$ \citep{Uskov24}.
\subsection*{ SRGA J025208.6+482952 }
A Seyfert 1.9 at $z=0.0337$ \citep{2023AstL...49...25U}.
\subsection*{ SRGA J025234.6+431001 }
A Seyfert 2 at $z=0.0512$ \citep{2022AstL...48...87U}.
\subsection*{ SRGA J025900.3+502958 }
Associated with LEDA\,2374943, a Seyfert 1 galaxy at $z=0.0946$ \citep{Uskov24}.
\subsection*{ SRGA J030536.3+762243 }
A Seyfert 1 at $z=0.1966$ \citep{2022ApJS..261....5M}.
\subsection*{ SRGA J030838.7-552042 }
Associated with LEDA\,410289, a Seyfert 2 galaxy at $z=0.0779$ \citep{2022ApJS..258...29C}.
\subsection*{ SRGA J031130.8-315249 }
A CV (nova-like and/or polar, \citealt{2003A&A...404..301R}).
\subsection*{ SRGA J031402.8+244442 }
Also a BL Lac candidate \citep{1998ApJS..118..127L}.
\subsection*{ SRGA J031415.5-542459 }
Possibly associated with WISEA\,J031416.45-542445.3, $W1-W2\sim 0.8$.
\subsection*{ SRGA J032317.9-481816 }
There are broad H$\beta$ and H$\gamma$ lines in the optical spectrum, $z=0.155$ \citep{2016AJ....152...25M}.
\subsection*{ SRGA J032511.9+404152 }
A pair of interacting Seyfert 2 galaxies, LEDA\,97012 and LEDA\,4678815 \citep{2010ATel.2759....1L,2012A&A...538A.123M}.
\subsection*{ SRGA J032718.6+552032 }
Unclassified radio-loud AGN \citep{2016MNRAS.460.3829M,2018ApJS..234....7M}.
\subsection*{ SRGA J032846.5-530316 }
Likely associated with the galaxy 2MASX\,J03284504-5303273, $W1-W2\sim 0.7$.
\subsection*{ SRGA J032910.2-220116 }
A Seyfert 1.8 \citep{2022ApJS..261....8R}.
\subsection*{ SRGA J033203.9-704000 }
A blazar at $z=0.277$ \citep{2021AJ....162..177P}.
\subsection*{ SRGA J033336.5-370656 }
The position of the ART-XC source is 44 arcsec away from the position of the Seyfert galaxy CTS 77, but another known Seyfert galaxy, FCC B868 ($z=0.0656$), located at 93 arcsec from the ART-XC source, may contribute to the X-ray signal.
\subsection*{ SRGA J033622.6-605854 }
A Seyfert 1 at $z=0.471$ \citep{2016AJ....152...25M}.
\subsection*{ SRGA J035023.8-501809 }
A Seyfert 2 \citep{2022ApJS..261....2K}.
\subsection*{ SRGA J035745.5+415457 }
A Seyfert 2 at $z=0.0518$ \citep{2022ApJS..261....2K}.
\subsection*{ SRGA J040111.1-535446 }
Associated with the gamma-ray source 4FGL\,J0401.0-5353, a blazar candidate of unknown type \citep{2020ApJ...892..105A}.
\subsection*{ SRGA J040335.6+472440 }
Associated with 2MASS\,J04033641+4724383, a Seyfert 1 galaxy at $z=0.0962$ \citep{Uskov24}.
\subsection*{ SRGA J040543.1+570730 }
An RS CVn variable \citep{2006A&A...454..301F}.
\subsection*{ SRGA J040551.3+380353 }
A flat-spectrum radio source \citep{2020ApJ...901....3I}.
\subsection*{ SRGA J040753.2-611607 }
A Seyfert 2 \citep{2022ApJS..258...29C}.
\subsection*{ SRGA J040850.6-791410 }
Likely associated with the star Gaia DR3 4625832751643853696 
at a distance $\sim 1300$\,pc.
\subsection*{ SRGA J041108.8-591127 }
A dwarf nova \citep{2023AJ....166..131T}.
\subsection*{ SRGA J041243.1+582520 }
A Seyfert 1 at $z=0.0684$ \citep{2022ApJS..261....5M}.
\subsection*{ SRGA J041328.1-061455 }
A flat-spectrum radio source \citep{2007ApJS..171...61H}.
\subsection*{ SRGA J041733.7-525302 }
Likely associated with the galaxy 2MASX\,J04173452-5253070, $W1-W2\sim 0.2$.
\subsection*{ SRGA J042435.7-754140 }
Likely associated with the red giant HD\,28839 at a distance $\sim 190$\,pc.
\subsection*{ SRGA J042616.8-592325 }
Associated with the galaxy LEDA\,371722, $W1-W2\sim 0.9$.
\subsection*{ SRGA J043008.4+455702 }
The bright, long-period variable V591 Per at a distance $\sim 1000$\,pc, a candidate binary with an orbital period of 2.48 days (TESS, \citealt{2023MNRAS.522...29G}).
\subsection*{ SRGA J043208.9+354922 }
A Seyfert 1 at $z=0.0506$ \citep{2021AstL...47...71Z}.
\subsection*{ SRGA J043510.7-752749 }
Likely associated with the variable star CRTS\,J043509.7-752743 at a distance $\sim 500$\,pc.
\subsection*{ SRGA J043516.4+512821 }
Based on the Swift/XRT position \citep{2023MNRAS.518..174E}, likely associated with WISEA\, J043519.72+512833.4, $W1-W2\sim 1.2$, also a radio source. A note of caution: the {\it Swift}/XRT position is 34 arcsec away from the ART-XC position,  i.e. significantly outside $R_{98}$.
\subsection*{ SRGA J043523.0+552235 }
An X-ray transient, SRGA J043520.9+552226 = SRGE J043523.3+552234, discovered by SRG/ART-XC and eROSITA, associated with the optical transient AT2019wey. An LMXB and a black hole candidate \citep{2021ApJ...920..120Y,2022A&A...661A..32M}.
\subsection*{ SRGA J043838.8-661405 }
Likely associated with the galaxy Fairall 304, $W1-W2\sim 0.5$.
\subsection*{ SRGA J043943.9-090312 }
A Seyfert 2 \citep{2022ApJS..258...29C}.
\subsection*{ SRGA J044048.9+292443 }
Associated with the star Gaia DR3 158134751604204416 at a distance of $\sim 300$ pc.
\subsection*{ SRGA J044956.8-642127 }
Possibly associated with WISEA\,J044956.29-642136.2, $W1-W2\sim 0.9$.
\subsection*{ SRGA J045001.1-494525 }
Possibly associated with WISEA\,J045001.39-494525.1, $W1-W2\sim 0.8$.
\subsection*{ SRGA J045001.8-551240 }
A Seyfert 2 at $z=0.0216$ \citep{2022ApJS..261....2K}.
\subsection*{ SRGA J045048.9+301443 }
A Seyfert 1.9 at $z=0.0331$ \citep{2021AstL...47...71Z}.
\subsection*{ SRGA J045253.8-585350 }
A blazar candidate \citep{2019ApJS..242....4D}.
\subsection*{ SRGA J045424.1-570045 }
Likely associated with the galaxy WISEA\,J045424.13-570045.9, $W1-W2\sim 0.8$, a radio source.
\subsection*{ SRGA J045431.4+524008 }
A Seyfert 1.9 at $z=0.0312$ \citep{2023AstL...49...25U}.
\subsection*{ SRGA J045442.2-431419 }
A Seyfert 1.9 \citep{2017ApJ...850...74K}.
\subsection*{ SRGA J045602.7+273600 }
An Orion variable.
\subsection*{ SRGA J050021.4+523801 }
Associated with the gamma-ray source 4FGL\,J0500.2+5237, likely a blazar \citep{2019A&A...632A..77C}.
\subsection*{ SRGA J050125.9-703338 }
In the LMC.
\subsection*{ SRGA J050249.5-622749 }
Associated with WISEA\,J050250.64-622740.6, $W1-W2\sim 1.1$.
\subsection*{ SRGA J050434.1-553129 }
Likely associated with the galaxy ESO 158-17, $W1-W2\sim 0.2$.
\subsection*{ SRGA J050809.7-660656 }
A new Be X-ray pulsar (eRASSU J050810.4-660653) discovered by SRG/eROSITA in the LMC \citep{2020ATel13609....1H,2022MNRAS.514.4018S}, previously detected during the ROSAT all-sky survey (2RXS J050810.2-660645).
\subsection*{ SRGA J050956.7-641743 }
A BL Lac at $z=0.271$ \citep{2021AJ....162..177P}.
\subsection*{ SRGA J051111.0-110305 }
A Seyfert 1 at $z=0.1168$ (Uskov et al., in prep.).
\subsection*{ SRGA J051300.0-252322 }
Possibly associated with the variable star \citep{2018MNRAS.477.3145J,2018AJ....156..241H} Gaia DR3 2956561001583884800, at a distance $\sim 600$ pc.
\subsection*{ SRGA J051313.8+662746 }
A Seyfert 2 at $z=0.0148 $ \citep{2023AstL...49...25U}.
\subsection*{ SRGA J051400.8-402723 }
Likely associated with the galaxy LEDA\,586143, $W1-W2\sim 0.8$, a radio source.
\subsection*{ SRGA J051401.8-504559 }
Likely associated with WISEA\,J051401.99-504603.5, $W1-W2\sim 1.0$, a radio source.
\subsection*{ SRGA J052028.6-715734 }
In the LMC.
\subsection*{ SRGA J052115.1+251331 }
A CV \citep{2020AJ....160....6T}.
\subsection*{ SRGA J052412.0-662049 }
In the LMC.
\subsection*{ SRGA J052503.0-693848 }
In the LMC.
\subsection*{ SRGA J052602.5-660510 }
In the LMC.
\subsection*{ SRGA J052627.5-211716 }
A Seyfert 2 \citep{2022ApJS..261....2K}.
\subsection*{ SRGA J052802.4-393451 }
A Seyfert 1 at $z=0.0367$ \citep{2022ApJS..261....2K}.
\subsection*{ SRGA J052858.5-670957 }
In the LMC.
\subsection*{ SRGA J052915.0-662447 }
A new Be X-ray Pulsar, eRASSU\,J052914.9-662446, discovered by SRG/eROSITA in the LMC \citep{2023A&A...669A..30M}.
\subsection*{ SRGA J052947.7-655645 }
In the LMC.
\subsection*{ SRGA J052959.2-340205 }
A Seyfert 1.8 \citep{2022AstL...48...87U}.
\subsection*{ SRGA J053013.2-655127 }
In the LMC.
\subsection*{ SRGA J053043.0-665431 }
In the LMC.
\subsection*{ SRGA J053232.2-655144 }
In the LMC.
\subsection*{ SRGA J053249.5-662217 }
In the LMC.
\subsection*{ SRGA J053322.0-684130 }
In the LMC.
\subsection*{ SRGA J053358.8-714528 }
Likely associated with the galaxy ESO 56-154, $W1-W2\sim 0.1$, a radio source.
\subsection*{ SRGA J053411.7-045030 }
There are two {\it Chandra} sources within the ART-XC localization region: 2CXO J053412.8$-$045035 and 2CXO J053410.4$-$045038, each with an X-ray flux $\sim 3\times 10^{-13}$~erg~s$^{-1}$~cm$^{-2}$ \citep{2010ApJS..189...37E}. Both are associated with Orion variable stars.
\subsection*{ SRGA J053526.1-691609 }
In the LMC.
\subsection*{ SRGA J053528.7-050541 }
Possibly associated with the Orion variable V1740 Ori, but other nearby stars may contribute to the flux.
\subsection*{ SRGA J053738.8+210827 }
A gamma Cas star.
\subsection*{ SRGA J053748.0-691021 }
In the LMC.
\subsection*{ SRGA J053856.8-640503 }
In the LMC.
\subsection*{ SRGA J053939.0-694439 }
In the LMC.
\subsection*{ SRGA J054011.1-691956 }
In the LMC.
\subsection*{ SRGA J054023.2-554443 }
Based on the {\it Swift}/XRT position (2SXPS\,J054022.5-554445), likely associated with the galaxy 2MASS\,J05402254-5544457, $W1-W2\sim 0.6$.
\subsection*{ SRGA J054134.6-682542 }
In the LMC.
\subsection*{ SRGA J055052.7-621453 }
A Seyfert 1 \citep{2022AstL...48...87U,2022ApJS..258...29C}.
\subsection*{ SRGA J055523.5-765506 }
Likely associated with the galaxy IC\,2160, $W1-W2\sim 0.1$, a radio source. Appears to be a Seyfert 2 based on the 6dF spectrum, but the H$\alpha$ region is missing.
\subsection*{ SRGA J055805.0-564634 }
Associated with 2MASX\,J05580592-5646361, a Seyfert 2 galaxy \citep{2022ApJS..258...29C}.
\subsection*{ SRGA J060041.7+000619 }
A Seyfert 1.9 \citep{2023A&A...671A.152M}.
\subsection*{ SRGA J060045.5-645602 }
Likely associated with the galaxy ESO 86-49, $W1-W2\sim 0.1$, a radio source.
\subsection*{ SRGA J060211.3-673802 }
Possibly asscoiated with WISEA J060211.85-673806.2, $W1-W2\sim 1.1$.
\subsection*{ SRGA J060241.7-595155 }
Associated with LEDA\,178859, $W1-W2\sim 0.7$, which appears to be a Seyfert 2 galaxy based on the 6dF spectrum.
\subsection*{ SRGA J060431.7-395014 }
Likely associated with the galaxy LEDA\,593504, $W1-W2\sim 0.5$.
\subsection*{ SRGA J060650.6-624544 }
Associated with the galaxy LEDA\,340165, $W1-W2\sim 0.8$.
\subsection*{ SRGA J060729.5-614832 }
A heavily obscured AGN \citep{2009MNRAS.398.1165G,2020MNRAS.497..229A} in an edge-on galaxy.
\subsection*{ SRGA J061035.2-652521 }
A Seyfert 1 \citep{2022ApJS..258...29C}.
\subsection*{ SRGA J061111.1-095612 }
Possibly associated with the star Gaia DR3 3004437272613544576, at a distance $\sim 700$ pc.
\subsection*{ SRGA J061154.6-435703 }
Likely associated with the galaxy LEDA\,542558, $W1-W2\sim 0.6$.
\subsection*{ SRGA J061324.1-290027 }
A Seyfert 2 at $z=0.0705$ \citep{2022AstL...48...87U}.
\subsection*{ SRGA J061448.3+093421 }
Likely associated with the variable star CY Ori, which has been tentatively classified as a dwarf nova\footnote{ \url{https://www.aavso.org/vsx/index.php?view=detail.top&oid=23188}}.
\subsection*{ SRGA J061619.0-705229 }
Based on the XMM-Newton position (4XMM\,J061617.1-705229, \citealt{2020A&A...641A.136W}), likely associated with the infrared source WISEA\,J061617.13-705228.7, $W1-W2\sim 0.4$.
\subsection*{ SRGA J061903.6-694744 }
Possibly associated with the galaxy 2MASX\,J06190057-6947399, $W1-W2\sim 0.0$.
\subsection*{ SRGA J062040.0+264338 }
The redshift is $z=0.134$ \citep{2021AJ....162..177P}.
\subsection*{ SRGA J062109.9-680551 }
Associated with the galaxy LEDA\,179145,  a Seyfert 2 \citep{2022ApJS..258...29C}.
\subsection*{ SRGA J062339.4-265803 }
Also known as SRGt\,062340.2-265751, a novalike CV \citep{2022A&A...661A..42S}.
\subsection*{ SRGA J062627.7+072726 }
A Seyfert 2 at $z=0.0425$ \citep{2022AstL...48...87U}.
\subsection*{ SRGA J062946.0-834423 }
Likely associated with WISEA\,J062948.62-834421.6, $W1-W2\sim 1.0$.
\subsection*{ SRGA J063326.9-561425 }
A Seyfert 2 \citep{2022AstL...48...87U}.
\subsection*{ SRGA J063517.1-695405 }
Associated with the galaxy WISEA\,J063517.18-695401.4 = 6dFGS\,gJ063517.2-695402, $W1-W2\sim 1.0$, a radio source.
\subsection*{ SRGA J063558.5+075525 }
Associated with the bright star TYC 733-2098-1 at a distance $\sim 200$ pc, not a CV (Zaznobin et al., in prep.).
\subsection*{ SRGA J064151.5-032039 }
A flat-spectrum radio quasar at $z=1.196$ \citep{2016ApJ...826...76A}.
\subsection*{ SRGA J064421.9-662623 }
Associated with the galaxy 6dFGS\,gJ064421.9-662620, $W1-W2\sim 0.8$, a Seyfert 1 galaxy based on the 6dF spectrum (Uskov et al., in prep.).
\subsection*{ SRGA J064620.5-692813 }
Associated with the galaxy 2MASX\,J06462178-6928111, a Seyfert 2 \citep{2022ApJS..258...29C}.
\subsection*{ SRGA J064849.8-694520 }
A blazar at $z=0.233$ \citep{2021AJ....162..177P}.
\subsection*{ SRGA J065018.7-380522 }
A narrow-line Seyfert 1 galaxy \citep{2018A&A...615A.167C}.
\subsection*{ SRGA J065313.0-673633 }
Associated with the dwarf nova ASASSN-V\,J065311.74-673642.9 \citep{2023A&A...675A.195S}.
\subsection*{ SRGA J065316.2-670832 }
Possibly associated with WISE\,J065313.86-670811.1, $W1-W2\sim 1.1$, a radio source.
\subsection*{ SRGA J065513.2-012841 }
Also known as MAXI J0655-013. A Be HMXB \citep{2022ATel15582....1Z,2022ATel15612....1R} and an X-ray pulsar \citep{2022ATel15495....1S}.
\subsection*{ SRGA J065638.8-670224 }
Possibly associated with the star Gaia DR3 5281660651288601472
at a distance $\sim 640$\,pc.
\subsection*{ SRGA J065721.8-670549 }
Possibly associated with the star Gaia DR3 5281647899528664320
at a distance $\sim 900$\,pc.
\subsection*{ SRGA J070110.6-323451 }
Associated with the galaxy 2MASS\,J07011023-3234529, $W1-W2\sim 0.8$.
\subsection*{ SRGA J070236.6-704441 }
Likely associated with the galaxy LEDA\,272172, $W1-W2\sim 0.1$.
\subsection*{ SRGA J070511.2-670531 }
Possibly associated with WISEA\,J070508.26-670525.9, $W1-W2\sim 1.4$.
\subsection*{ SRGA J070637.0+635109 }
A Seyfert 1.8 at $z=0.0140$  \citep{2022AstL...48...87U}.
\subsection*{ SRGA J070932.6-353737 }
A Seyfert 1.5 at $z=0.030$ \citep{2017A&A...602A.124R}.
\subsection*{ SRGA J071020.4-241648 }
Likely associated with the galaxy 2MASX\,J07102131-2416501, $W1-W2\sim 0.5$, a radio source.
\subsection*{ SRGA J071029.2-390131 }
Appears to be a Seyfert 1 galaxy based on the 6dF spectrum.
\subsection*{ SRGA J071415.0-262116 }
Associated with the Be star 27 CMa at $\sim 440$ pc. Possibly, a gamma Cas star.
\subsection*{ SRGA J071740.2-710346 }
Associated with WISEA\,J071740.40-710347.3, $W1-W2\sim 1.0$.
\subsection*{ SRGA J071947.9-753801 }
Associated with the galaxy 2MASS\,J07195059-7537573, $W1-W2\sim 0.7$.
\subsection*{ SRGA J072041.0-552615 }
Associated with the galaxy LEDA\,409410, $W1-W2\sim 0.6$.
\subsection*{ SRGA J072319.9-732653 }
A cluster of galaxies discovered via the Sunyaev-Zeldovich effect by the {\it Planck observatory} and confirmed in X-rays by {\it XMM-Newton} \citep{2011A&A...536A...9P}.
\subsection*{ SRGA J072957.0-654331 }
A Seyfert 1 galaxy \citep{2022ApJS..258...29C}.
\subsection*{ SRGA J073022.2-750414 }
Likely associated with 2MASS\,J07302521-7504018, $W1-W2\sim 1.0$.
\subsection*{ SRGA J074343.1+183157 }
Possibly associated with the galaxy WISEA\,J074343.10+183143.5, $W1-W2\sim 0.2$.
\subsection*{ SRGA J074414.2-704131 }
Likely associated with 2MASS\,J07441473-7041302, $W1-W2\sim 0.8$, a radio source.
\subsection*{ SRGA J075808.4+835632 }
A Seyfert 1 at $z=0.1340$ \citep{2022ApJS..261....2K}.
\subsection*{ SRGA J080142.6+420013 }
The known AGN SDSS\,J080142.58+420019.4, previously undetected in X-rays.
\subsection*{ SRGA J080559.6+320605 }
Likely associated with the galaxy 2MASS\,J08055841+3205542, $W1-W2\sim 1.2$.
\subsection*{ SRGA J080917.6+474637 }
Possibly associated with the extended optical object WISEA\,J080917.45+474651.8, $W1-W2\sim 0.5$.
\subsection*{ SRGA J082316.0-632939 }
The redshift is $z=0.29$ \citep{2022ApJS..263...24A}.
\subsection*{ SRGA J082623.5-703144 }
Also known as PBC\,J0826.3-7033. Likely a non-magnetic CV \citep{2012A&A...545A.101P}.
\subsection*{ SRGA J083126.1-600717 }
Likely associated with the star Gaia DR3 5302213306753527936, at  a roughly estimated distance $\sim 7.5$ kpc.
\subsection*{ SRGA J084433.5-375751 }
Associated with the star HD\,74771, a bright variable star \citep{2017ARep...61...80S}.
\subsection*{ SRGA J084551.9-272922 }
Possibly associated with the star Gaia DR3 5645994171436695552, at a distance $\sim 600$ pc.
\subsection*{ SRGA J084937.3-554416 }
Likely associated with one of two stars: Gaia DR3 5316706343179643904 or Gaia DR3 5316706347473511936, at distances $\sim 1.5$ and $\sim 3$ kpc, respectively.
\subsection*{ SRGA J085040.9-421155 }
A red-supergiant HMXB \citep{2024MNRAS.528L..38D}.
\subsection*{ SRGA J085741.3-554226 }
Possibly associated with WISEA\,J085739.32-554258.2, $W1-W2\sim 2.1$.
\subsection*{ SRGA J090106.7-340025 }
Likely associated with the galaxy WISEA\,J090108.25-340024.3, $W1-W2\sim 1.0$, a radio source.
\subsection*{ SRGA J090305.2+130323 }
Possibly associated with the star Gaia DR3 605340838251799296, at a distance $\sim 1200$ pc.
\subsection*{ SRGA J091514.0-752345 }
Associated with the galaxy 2MASS\,J09151520-7523498, $W1-W2\sim 1.1$, a radio source.
\subsection*{ SRGA J092014.4-383445 }
Likely associated with WISEA\,J092014.24-383450.4 , $W1-W2\sim 0.7$, a strong radio source.
\subsection*{ SRGA J092418.1-314217 }
A LMXB or CV \citep{2017ApJS..230...25T}.
\subsection*{ SRGA J092712.2-113828 }
A Seyfert 2 at $z=0.0109$ (Uskov et al., in prep.).
\subsection*{ SRGA J092841.6-620754 }
Possibly associated with WISEA\,J092841.59-620741.4, $W1-W2\sim 1.2$.
\subsection*{ SRGA J095307.9-765751 }
A blazar at $z=0.109$ \citep{2021AJ....162..177P}.
\subsection*{ SRGA J095704.9-585317 }
Possibly associated the radio/IR source RACS-DR1\,J095702.6-585317 = CatWISE2020\,J095702.71-585318.5, $W1-W2\sim 1.1$.
\subsection*{ SRGA J100513.7-625209 }
A Seyfert 1 at $z=0.0714$ \citep{2022ApJS..261....5M}.
\subsection*{ SRGA J101133.2-442257 }
Based on the Swift/XRT position, possibly associated with the radio/IR source RACS-DR1\,J101132.0-442253 = WISEA\,J101132.10-442253.5, $W1-W2\sim 0.4$.
\subsection*{ SRGA J102205.8-353801 }
Likely associated with the hot subdwarf candidate GALEX\,J102205.1-353756 at a distance $\sim 900$ pc.
\subsection*{ SRGA J102555.0-574844 }
A very massive, colliding-wind binary.
\subsection*{ SRGA J103915.5-490307 }
Associated with the star Gaia DR3 5364403368050904576, at a distance $\sim ~3700$ pc.
\subsection*{ SRGA J104124.3-692952 }
Possibly associated with the galaxy 2MASX\,J10412355-6929560, $W1-W2\sim 0.4$.
\subsection*{ SRGA J104451.4-602510 }
A Seyfert 2 at $z=0.0470$ \citep{2022ApJS..261....2K}.
\subsection*{ SRGA J104834.8-390225 }
A Seyfert 1.2 at $z=0.0449$ \citep{2022ApJS..261....5M}.
\subsection*{ SRGA J105452.7+771252 }
Possibly associated with WISEA J105452.40+771310.0, $W1-W2\sim 0.7$.
\subsection*{ SRGA J105715.0-473958 }
A Seyfert 1.9 \citep{2018ApJS..235....4O}.
\subsection*{ SRGA J110122.2-514909 }
Associated with the galaxy 2MASX\,J11012138-5149100 , $W1-W2\sim 0.8$.
\subsection*{ SRGA J110946.0+800817 }
A Seyfert 1 at $z=0.1888$ \citep{2023AstL...49...25U}.
\subsection*{ SRGA J111457.3-611449 }
A massive star-forming region. {\it Chandra} has registered plenty of point X-ray sources and diffuse X-ray emission \citep{2014ApJS..213....1T}.
\subsection*{ SRGA J111515.8-480620 }
Possibly associated with WISEA\,J111515.15-480615.4, $W1-W2\sim 1.0$.
\subsection*{ SRGA J111820.6-543730 }
Likely a LMXB, rather than an HMXB \citep{2013A&A...560A.108C}.
\subsection*{ SRGA J112958.5-655520 }
Based on the {\it Swift}/XRT coordinates, likely associated with the star Gaia DR3 5236824109000257920, a long-period variable according to Gaia. The distance is at least several kpc, according to Gaia.
\subsection*{ SRGA J113151.7-123151 }
An AGN at $z=0.658$ lensed by a galaxy at $z=0.295$ \citep{2003A&A...406L..43S}.
\subsection*{ SRGA J113257.1-260736 }
Associated with the galaxy WISEA\,J113256.57-260735.5, $W1-W2\sim 1.3$, a radio source.
\subsection*{ SRGA J113850.8-232126 }
A Seyfert 1.9 \citep{2022ApJS..261....2K}.
\subsection*{ SRGA J114515.2+794100 }
A Seyfert 1 at $z=0.0060$ \citep{2022ApJS..261....2K}.
\subsection*{ SRGA J114722.1-495308 }
Possibly associated with the emission-line star TWA 19B, but there is also another bright star, HD 102458, in the vicinity.
\subsection*{ SRGA J114754.6+094554 }
An AGN in a galaxy pair \citep{2011ApJ...737..101L}.
\subsection*{ SRGA J115215.3-510703 }
Possibly associated with the star Gaia DR3 5369277537456720128 at a distance $\sim 660$\,pc,
\subsection*{ SRGA J115223.8-673837 }
Likely associated with the star Gaia DR3 5235694124616931584 at a distance $\sim 1050$\,pc.
\subsection*{ SRGA J115415.6-501801 }
Possibly associated with the star Gaia DR3 5370642890382757888 at a distance of $\sim 310$\,pc.
\subsection*{ SRGA J115730.2-343758 }
Associated with the CV 6dFGS\,g1157297-343750 \citep{2010MNRAS.401.1151M}.
\subsection*{ SRGA J115910.3-532435 }
Likely associated with the galaxy ESO 171-4, $W1-W2\sim 0.1$, a radio source \citep{2007MNRAS.382..382M}.
\subsection*{ SRGA J120413.5-294651 }
Possibly associated with WISEA\,J120412.71-294709.1, $W1-W2\sim 0.9$, a radio source.
\subsection*{ SRGA J121053.2-040811 }
Possibly associated with the star Gaia DR3 3597896582057220096 at a distance $\sim 900$ pc.
\subsection*{ SRGA J122806.9-552242 }
Likely associated with the rotating variable star ASAS\,J122810-5523.0 at a distance $\sim 110$\,pc.
\subsection*{ SRGA J122809.9-092718 }
A Seyfert 1.5 at $z=0.223$ \citep{2017A&A...602A.124R}.
\subsection*{ SRGA J123112.7-423529 }
Likely associated with the star Gaia DR3 6145666951501477376 at a distance $\sim 400$\,pc.
\subsection*{ SRGA J123402.7-614310 }
A transient hard X-ray source, discovered by {\it INTEGRAL} and confirmed by {\it Swift}/XRT, likely a supergiant fast X-ray transient \citep{2020MNRAS.491.4543S,2020ATel14039....1S}.
\subsection*{ SRGA J123630.6-664551 }
Likely associated with the star Gaia DR3 5859908625444588160 at a distance $\sim 400$\,pc.
\subsection*{ SRGA J123821.5-253208 }
A bright, short-duration X-ray transient, SRGt\,J123822.3-253206, discovered by ART-XC and eROSITA \citep{2020ATel13415....1S,2020ATel13416....1W}.
\subsection*{ SRGA J124248.7-630324 }
A gamma Cas star \citep{2018PASJ...70..109T,2020A&A...633A..40L}.
\subsection*{ SRGA J124403.4-632220 }
An X-ray pulsar with a Be companion, discovered by SRG/ART-XC and eROSITA, also known as SRGA\,J124404.1-632232 and SRGU\,J124403.8-632231 \citep{2022A&A...661A..21D}.
\subsection*{ SRGA J130522.0-492823 }
In NGC 4945.
\subsection*{ SRGA J131036.8-562654 }
A Seyfert 2 at $z=0.1142$.
\subsection*{ SRGA J131239.9-624258 }
A Wolf-Rayet star. The most likely physical picture is that of colliding stellar winds in a wide binary system, with the unseen secondary star being another WR star or a luminous blue variable \citep{2014ApJ...785....8Z}.
\subsection*{ SRGA J132031.7-701443 }
Associated with the bright star Gaia DR3 5844075864125790464 at a distance $\sim 2300$\,pc. Possibly an HMXB or a symbiotic X-ray binary.
\subsection*{ SRGA J132327.2-165810 }
Associated with the galaxy LEDA\,158848, $W1-W2\sim 0.3$, a radio source.
\subsection*{ SRGA J132532.0-621312 }
Based on the XMM-Newton position (4XMM\,J132531.2-621309, \citealt{2020A&A...641A.136W}), likely associated with the near-IR source VVV\,358613430	\citep{2018MNRAS.474.1826S}.
\subsection*{ SRGA J133836.9-352851 }
Associated with the galaxy 2MASS\,J13383728-3528517, $W1-W2\sim 1.3$.
\subsection*{ SRGA J133950.6-643002 }
Possibly associated with IGR\,J13402-6428 at 2.8 arcmin from the ART-XC position. However, the latter is inconsistent with the positions of two {\it Chandra} sources, CXOU\,J133935.8-642537 and CXOU\,J133959.2-642444, which have been suggested as possible soft X-ray counterparts of IGR\,J13402-6428 \citep{2012ApJ...754..145T}.
\subsection*{ SRGA J134742.7-621356 }
Likely, a supernova remnant \citep{2012ATel.3963....1R}.
\subsection*{ SRGA J135418.5-374648 }
The ART-XC source is 39\arcsec\ away from the optical position of the Seyfert 2 galaxy Tol 1351-375 and appears to consist of two sources.
\subsection*{ SRGA J135534.4+352045 }
A Seyfert 2 at $z=0.1021$ \citep{2022ApJS..261....2K}.
\subsection*{ SRGA J140130.2-493239 }
An FR II radio galaxy \citep{2006AJ....131..100B}.
\subsection*{ SRGA J140146.3-501333 }
Possibly associated with the star Gaia DR3 6090855029847914368, at a distance $\sim 1200$ pc.
\subsection*{ SRGA J140503.9-534130 }
Likely associated with the galaxy 2MASX\,J14050398-5341278, $W1-W2\sim 0.2$.
\subsection*{ SRGA J141249.4-402138 }
A CV of VY Scl-type \citep{2010AN....331..227G}.
\subsection*{ SRGA J142006.2-165410 }
Likely associated with 2MASS\,J14200653-1653573 = NVSS\,J142006-165357 = CatWISE\,J142006.51-165357.0, $W1-W2=1.2$, a flat-spectrum radio source \citep{2020ApJ...901....3I}.
\subsection*{ SRGA J142543.6-310716 }
Possibly associated with the galaxy 6dF\,J1425443-310721, $W1-W2\sim 0.5$.
\subsection*{ SRGA J143701.5+073508 }
Associated with SDSS\,J143701.26+073508.3, a Seyfert 2 galaxy at $z=0.18305$ (based on visual inspection of the SDSS spectrum), $W1-W2\sim 0.7$, a radio source.
\subsection*{ SRGA J143739.3-361323 }
The hot subdwarf OB+ candidate \citep{2019A&A...621A..38G} CD-359665 at a distance $\sim 600$\,pc.
\subsection*{ SRGA J144720.0-581607 }
Possibly associated with the luminous star Gaia DR3 5879166120570299776 at a distance $\sim 3200$\,pc.
\subsection*{ SRGA J144914.1-553620 }
A Seyfert 2 at $z=0.0187$.
\subsection*{ SRGA J145524.8-645953 }
Likely associated with WISEA\,J145523.78-650002.5, $W1-W2\sim 1.1$, which seems to be extended in DECaPS images.
\subsection*{ SRGA J145532.6-544636 }
Associated with the galaxy 2MASX\,J14553197-5446291, $W1-W2\sim 0.8$.
\subsection*{ SRGA J145738.5-362411 }
Possibly associated with WISEA\,J145737.30-362412.9, $W1-W2\sim 0.9$.
\subsection*{ SRGA J150407.2-024810 }
There is a weak central AGN, but the X-ray emission from the intracluster gas clearly dominates \citep{2005ApJ...633..148B}.
\subsection*{ SRGA J151338.1-344046 }
Associated with the changing-look Seyfert galaxy 2MASX\,J15133881-3440408 \citep{2022MNRAS.511...54H}.
\subsection*{ SRGA J151425.6-545900 }
Likely associated with the bright star Gaia DR3 5886714375305398656, at a distance $\sim 2$ kpc, which is an ellipsoidal binary candidate (TESS, \citealt{2023MNRAS.522...29G}).
\subsection*{ SRGA J151931.6-472604 }
Likely associated with the edge-on galaxy LEDA\,2793198, $W1-W2\sim 0.2$.
\subsection*{ SRGA J152102.1+320410 }
A Seyfert 2 at $z=0.1143$ \citep{2021AstL...47...71Z}.
\subsection*{ SRGA J153158.5-293020 }
A blazar candidate \citep{2019ApJS..242....4D}.
\subsection*{ SRGA J153309.3-522318 }
Possibly associated with the star Gaia DR3 5888715417783989760, at a distance $\sim 850$\,pc, a hot subdwarf candidate \citep{2019A&A...621A..38G}.
\subsection*{ SRGA J153412.6+625855 }
A Seyfert 1 at $z=0.2372$ \citep{2022ApJS..261....5M}.
\subsection*{ SRGA J153603.0-574849 }
Associated with WISEA\,J153602.80-574853.3, $W1-W2\sim 0.8$.
\subsection*{ SRGA J153814.2-554215 }
Likely a LMXB \citep{2012A&A...540A..22D}.
\subsection*{ SRGA J154426.3-201636 }
Associated wtih WISEA\,J154426.00-201634.3, $W1-W2\sim 1.1$.
\subsection*{ SRGA J154743.0-442203 }
Likely associated with WISEA\,J154743.17-442158.4, $W1-W2\sim 0.6$.
\subsection*{ SRGA J155432.1-334006 }
Likely associated with 2MASS\,J15543155-3340151, $W1-W2\sim 1.2$.
\subsection*{ SRGA J155603.4-521610 }
The non-magnetic CV ASASSN-19rc \citep{2023AJ....165..163C}.
\subsection*{ SRGA J155901.6-733759 }
Associated with WISEA\,J155902.56-733748.5, $W1-W2\sim 0.8$.
\subsection*{ SRGA J160050.8-514300 }
A colliding-wind Wolf-Rayet binary \citep{2019NatAs...3...82C}.
\subsection*{ SRGA J160455.4-722323 }
Associated with the flaring M-dwarf ASASSN-16hl  \citep{2019ApJ...876..115S}.
\subsection*{ SRGA J160534.2+164904 }
Associated with the known Seyfert 2 galaxy 2MASX\,J16053413+1649078 at $z=0.141$, previously undetected in X-rays.
\subsection*{ SRGA J160947.4-545110 }
Possibly associated with WISEA\,J160947.26-545107.8, $W1-W2\sim 0.6$.
\subsection*{ SRGA J161018.6-634242 }
A Seyfert 1.5, at $z=0.2094$ \citep{2018ATel11341....1S}.
\subsection*{ SRGA J161251.4-052105 }
A Seyfert 2 at $z=0.0306$ \citep{2023AstL...49...25U}.
\subsection*{ SRGA J161537.9+594546 }
Possibly associated with Gaia DR3 1624769330460296192, $W1-W2\sim 0.8$. which can be a galaxy. However, there is also a star, Gaia DR3 1624769330461204096, within 2 arcsec.
\subsection*{ SRGA J161728.9-502245 }
Possibly a CV \citep{2012ApJ...754..145T}.
\subsection*{ SRGA J161735.5+501447 }
The known AGN SBS\,1616+503, previously undetected in X-rays.
\subsection*{ SRGA J161944.1-132618 }
A Seyfert 1.9 at $z=0.0789$ \citep{2023AstL...49...25U}.
\subsection*{ SRGA J161947.6+435849 }
Possibly associated with the quasar SDSS\,J161946.29+435915.4, $z=0.8524$, a giant radio quasar \citep{2021ApJS..253...25K}. However, the offset (29.5 arcsec) is significantly larger than the error radius ($R_{98}=19.9$\,arcsec).
\subsection*{ SRGA J162312.1-260324 }
Possibly associated with WISEA\,J162311.77-260338.1, $W1-W2\sim 1.2$, and with the strong radio source NVSS\,J162312-260320.
\subsection*{ SRGA J162420.7-331101 }
A Seyfert 2 at $z=0.0279$ \citep{2022ApJS..261....6K}.
\subsection*{ SRGA J162618.5-392702 }
The known optical transient ASASSN-17ep \citep{2019MNRAS.486.1907J}, associated with either the star Gaia DR3 6017984243894659840 or with the star Gaia DR3 6017984243877931008, which are within 1 arcsec of each other.
\subsection*{ SRGA J163529.1-480606 }
A bright star at a distance $\sim 1100$ pc. Possibly an HMXB or a symbiotic X-ray binary.
\subsection*{ SRGA J163827.9-441447 }
Possibly associated with the bright radio source AT20G\,J163827-441444.
\subsection*{ SRGA J164046.2-032813 }
Possibly associated with the star Gaia DR3 4354844848120561920 at a distance $\sim 1200$ pc.
\subsection*{ SRGA J164050.9+654718 }
Associated with the edg-on galaxy UGC\,10519, an obscured AGN at $z=0.0244$ (Uskov et al., in prep.).
\subsection*{ SRGA J164356.8-321347 }
Possibly associated with the infrared and radio source WISEA\,J164356.93-321404.1 = NVSS\,J164356-321409, $W1-W2\sim 0.7$.
\subsection*{ SRGA J164433.4-513411 }
Possibly associated with WISEA\,J164433.25-513413.2, $W1-W2\sim 0.3$ (CatWISE), which appears to be extended in DECaPS images.
\subsection*{ SRGA J165143.2+532539 }
A known X-ray source, 2E\,1650.6+5330 = 2RXS\,J165144.7+532532, associated with SBS\,1650+535, a Seyfert 1.8 galaxy at $z=0.0286$ \citep{Uskov24}.
\subsection*{ SRGA J165251.2+172700 }
Based on the XMM-Newton coordinates (4XMM\,J165251.4+172651, \citealt{2020A&A...641A.136W}), associated with the galaxy 2MASX\,J16525143+1726514, $W1-W2\sim 0.2$.
\subsection*{ SRGA J165552.1-495737 }
A Seyfert 2 at $z=0.0586$ \citep{2023A&A...671A.152M}.
\subsection*{ SRGA J170025.6-724043 }
A Seyfert 1 \citep{2022ApJS..258...29C}.
\subsection*{ SRGA J170336.8+620130 }
An ambiguous nuclear transient (has some characteristics seen in both tidal disruption events and active galactic nuclei, \citealt{2022ApJ...930...12H}) in the galaxy NGC 6297.
\subsection*{ SRGA J170412.5-443136 }
Also known as XTE J1704-445, a strongly variable X-ray source \citep{2007ATel.1164....1M}.
\subsection*{ SRGA J170723.3-194422 }
Based on the {\it XMM-Newton} coordinates (XMMSL2\,J170722.0-194429), likely associated with WISEA\,J170722.22-194426.3, $W1-W2\sim 0.3$, a radio source.
\subsection*{ SRGA J171115.2+530953 }
Likely associated with WISEA\,J171115.73+530948.1, $W1-W2=1.0$.
\subsection*{ SRGA J171334.7-252015 }
Likely associated with WISEA\,J171334.96-252015.4, $W1-W2\sim 0.7$.
\subsection*{ SRGA J171544.9-232608 }
Possibly associated with the star Gaia DR3 4114267164182444032, at a distance $\sim 680$~pc, $W1-W2\sim 0.5$, a radio source.
\subsection*{ SRGA J173253.5-440735 }
Likely associated with the star Gaia DR3 5958345771178324224, at a distance $\sim 1650$~pc.
\subsection*{ SRGA J173338.1+363133 }
A Seyfert 1.9 at $z=0.0437$ \citep{2022ApJS..261....2K}.
\subsection*{ SRGA J173918.4-545409 }
Possibly associated with WISEA\,J173918.41-545401.2, $W1-W2\sim 0.9$.
\subsection*{ SRGA J174008.8-284721 }
Likely an intermediate polar rather than a LMXB \citep{2013ApJ...769..120B,2014ApJ...786...20L}.
\subsection*{ SRGA J174027.1-365542 }
Intermediate polar \citep{2019ApJ...887...32C}.
\subsection*{ SRGA J174038.1-273658 }
A new X-ray transient discovered by Swift/XRT on 9 April 2021 \citep{2021ATel14536....1B}.
\subsection*{ SRGA J174046.3+060353 }
A CV \citep{2015AJ....150..170H}.
\subsection*{ SRGA J174201.3-605516 }
A Seyfert 1.5 at $z=0.152$ \citep{2017A&A...602A.124R}, a FR II radio galaxy \citep{2022ApJS..262...51M}.
\subsection*{ SRGA J174242.0+184112 }
A close binary system \citep{2018AJ....156..241H} at a distance $\sim 200$\,pc, with an X-ray luminosity $\sim 2\times 10^{31}$\,erg\,s$^{-1}$. Hence, possibly an RS CVn type system.
\subsection*{ SRGA J174445.6-295046 }
A symbiotic X-ray binary \citep{2014MNRAS.441..640B}.
\subsection*{ SRGA J174850.6+665403 }
Associated with 2MASX\,J17485018+6653544, a Seyfert 2 galaxy at $z=0.0261$ (Uskov et al., in prep.).
\subsection*{ SRGA J175328.6-244632 }
A gamma Cas star.
\subsection*{ SRGA J175340.5+654239 }
A Seyfert 1.9 at $z=0.1487$ (Uskov et al., in prep.).
\subsection*{ SRGA J175721.0-304409 }
Likely a LMXB with a giant companion \citep{2013AdSpR..51.1278M}.
\subsection*{ SRGA J175758.4+042732 }
Associated with the star Gaia DR3 4472788803304480896, a CV  (Zaznobin et al., in prep.).
\subsection*{ SRGA J175806.5+660038 }
Associated with LEDA\,2682746, a Seyfert 1.9 galaxy at $z=0.0871$ (Uskov et al., in prep.).
\subsection*{ SRGA J180312.5+045110 }
Likely associated with the infrared source WISEA\,J180313.39+045115.4, $W1-W2\sim 0.9$.
\subsection*{ SRGA J180342.0+615651 }
Associated with the infrared and radio source WISEA\,J180341.28+615653.3, a Seyfert 1 galaxy at $z=0.4178$ (Uskov et al., in prep.).
\subsection*{ SRGA J180639.3+663247 }
Associated with 2MASS\,J18063992+6632411, a Seyfert 2 galaxy at $z=0.0867$ (Uskov et al., in prep.).
\subsection*{ SRGA J180731.6+662950 }
Associated with WISEA\,J180733.80+662941.8, a Seyfert 1.9 galaxy at $z=0.2615$ (Uskov et al., in prep.).
\subsection*{ SRGA J180849.4+663426 }
A flat-spectrum radio source \citep{2014ApJS..213....3M}.
\subsection*{ SRGA J181227.8-181237 }
An ultra-compact LMXB and burster \citep{2019MNRAS.486.4149G}.
\subsection*{ SRGA J181239.8-221924 }
An X-ray transient. Likely a black-hole X-ray binary, based on X-ray and radio properties \citep{2022MNRAS.513.6196R}.
\subsection*{ SRGA J181636.0-391251 }
Likely an intermediate polar associated with the star Gaia DR3 6727363681257990016 \citep{2020AstL...45..836K}.
\subsection*{ SRGA J181749.5+234311 }
Associated with LEDA\,1692433, a Seyfert 1.9 galaxy at $z=0.0813$ \citep{Uskov24}.
\subsection*{ SRGA J181815.4-545033 }
Possibly associated with WISEA\,J181814.61-545027.2, $W1-W2\sim 0.7$.
\subsection*{ SRGA J182111.0+765816 }
A Seyfert 2 at $z=0.0631$ \citep{2023AstL...49...25U}.
\subsection*{ SRGA J182156.3+642033 }
A powerful AGN in the central galaxy of a rich cluster. The X-ray luminosities of the quasar and cluster are comparable, according to previous observations, in particular Chandra \citep{2010MNRAS.402.1561R,2014ApJ...792L..41R}.
\subsection*{ SRGA J182510.7+645021 }
A KO giant star with blended double lines in the optical spectrum \citep{2019A&A...626A..31S}, thus most likely a coronally active binary.
\subsection*{ SRGA J182528.9+720905 }
A Seyfert 2 at $z=0.1108$ \citep{2022ApJS..261....2K}.
\subsection*{ SRGA J182832.6-592055 }
A blazar candidate \citep{2019A&A...632A..77C}.
\subsection*{ SRGA J183754.5+155436 }
Possibly associated with the bright variable star ASAS\,J183754+1554.7 = Gaia DR3 4510001297611019392 at a distance $\sim 500$~pc,
\subsection*{ SRGA J183811.5+655457 }
A Seyfert 1 at $z=0.2301$ (Uskov et al., in prep.).
\subsection*{ SRGA J183906.5-571501 }
Associated with the infrared source WISEA\,J183905.95-571505.1, $W1-W2\sim 1.8$; suggested to be a BL Lac based on the featureless optical spectrum \citep{2023A&A...671A.152M}. However, the optical counterpart (Gaia DR3 6637527637032850176) appears to have a large proper motion.
\subsection*{ SRGA J184006.8+592011 }
Likely associated with the galaxy NGC\,6696, $W1-W2\sim 0.3$.
\subsection*{ SRGA J184119.5-045617 }
The magnetar 1E 1841-045 in the supernova remnant Kes 73, which are both bright in X-rays \citep{2013ApJ...779..163A,2014ApJ...781...41K} and consistent with the ART-XC position.
\subsection*{ SRGA J185421.6+083842 }
Based on the position of the soft X-ray counterpart determined with {\it Swift}/XRT \citep{2018AstL...44..522K}, the object is likely associated with the infrared source CatWISE J185422.29+083846.1, $W1-W2\sim 1.0$, which is likely a star  (Zaznobin et al., in prep.).
\subsection*{ SRGA J190140.6+012628 }
An LMXB \citep{2012AstL...38..629K}.
\subsection*{ SRGA J190308.3+733241 }
Likely associated with the galaxy WISEA\,J190306.44+733246.0, $W1-W2\sim 0.8$.
\subsection*{ SRGA J190529.2-263916 }
Likely associated with WISEA\,J190528.57-263915.0, $W1-W2\sim 1.1$, a radio source.
\subsection*{ SRGA J190629.4-044655 }
Possibly associated with WISEA\,J190628.35-044655.2, $W1-W2\sim 0.7$.
\subsection*{ SRGA J190722.1-204641 }
Likely an intermediate polar \citep{2019MNRAS.489.3031X}.
\subsection*{ SRGA J191438.9+502846 }
Associated with the galaxy WISEA\,J191437.56+502854.6, $W1-W2\sim 0.9$, a radio source.
\subsection*{ SRGA J191456.9+103641 }
Likely a HMXB \citep{2015MNRAS.446.1041C}.
\subsection*{ SRGA J191628.1+711619 }
Associated with WISEA\,J191629.25+711616.4, a Seyfert 1 galaxy at $z=0.0984$ \citep{Uskov24}.
\subsection*{ SRGA J192413.0+665631 }
Likely associated with WISEA\,J192408.23+665624.1, $W1-W2\sim 0.8$, a radio source.
\subsection*{ SRGA J192501.9+504316 }
A Seyfert 1.2 at $z=0.068$ \citep{2013A&A...556A.120M}.
\subsection*{ SRGA J193346.0+325424 }
A Seyfert 1.2 at $z=0.063$ \citep{2007ApJ...669..109L}.
\subsection*{ SRGA J193707.1+660811 }
A narrow-line Seyfert 1 galaxy at $z=0.0714$ \citep{2023AstL...49...25U}.
\subsection*{ SRGA J194412.5-243619 }
Associated with 2MASX\,J19441243-2436217, a Seyfert 2 galaxy at $z=0.1402$ \citep{Uskov24}.
\subsection*{ SRGA J194438.6-043223 }
Possibly associated with the star Gaia DR3 4209574176412222336, at a distance $\sim 1200$\,pc.
\subsection*{ SRGA J194638.0+704554 }
An intermediate polar \citep{2022A&A...661A..39Z}.
\subsection*{ SRGA J195226.6+380011 }
Associated with 2MASS\,J19522509+3800269, a Seyfert 1 galaxy at $z=0.0767$ \citep{Uskov24}.
\subsection*{ SRGA J195702.7+615028 }
A Seyfert 1 at $z=0.0586$ \citep{2022AstL...48...87U}.
\subsection*{ SRGA J195815.6+194141 }
A Seyfert 1 at $z=0.0310$ \citep{2016AstL...42..295B}.
\subsection*{ SRGA J195927.9+404358 }
A significant contribution of the central AGN to the X-ray luminosity of the cluster is expected \citep{2002ApJ...564..176Y}.
\subsection*{ SRGA J200206.8+413240 }
Associated with the strong, flat-spectrum \citep{2018MNRAS.474.5008D} radio source NVSS\,J200206+413243.
\subsection*{ SRGA J200331.8+701331 }
A Seyfert 1 at $z=0.0976$ \citep{2023AstL...49...25U}.
\subsection*{ SRGA J200431.2+610218 }
A Seyfert 2 at $z=0.05866$ \citep{2021AstL...47...71Z}.
\subsection*{ SRGA J201116.5+600418 }
A CV (Zaznobin et al., in prep.).
\subsection*{ SRGA J201633.2+705525 }
Associated with WISEA\,J201632.61+705527.2, a Seyfert 1 galaxy at $z=0.2579$ \citep{Uskov24}.
\subsection*{ SRGA J201921.1+300257 }
Likely associated with the star Gaia DR3 1861455769748627584, at a distance $\sim 900$ pc, which is an ellipsoidal binary candidate (TESS, \citealt{2023MNRAS.522...29G}).
\subsection*{ SRGA J202835.2+254401 }
A Compton-thick AGN, MCG\,+04-48-002, at $z=0.0136$ in a galaxy pair with another Compton-thick AGN, NGC\,6921, at $z=0.0147$. The ART-XC source is associated with MCG\,+04-48-002, since NGC\,6921 is 91'' away \citep{2016ApJ...824L...4K}.
\subsection*{ SRGA J202931.7-614911 }
A Seyfert 2 at $z=0.1243$ \citep{2022ApJS..261....2K}.
\subsection*{ SRGA J203515.3+220628 }
Likely associated with the star Gaia DR3 1818529770639712384 at a distance $\sim 800$~pc. Suggested to be a pulsating hot subdwarf based on TESS photometry \citep{2022A&A...663A..45K}.
\subsection*{ SRGA J204149.7-373346 }
A blazar candidate \citep{2019A&A...632A..77C} or a cluster of galaxies \citep{2000AN....321....1S}.
\subsection*{ SRGA J204318.6+443820 }
An X-ray pulsar in a Be HMXB system, discovered by SRG/ART-XC and eROSITA \citep{2022A&A...661A..28L}, also known as SRGA\,J204318.2+443815 = SRGe\,J204319.0+443820.
\subsection*{ SRGA J204547.9+672640 }
A magnetic CV \citep{2022A&A...661A..39Z}.
\subsection*{ SRGA J205522.2+401808 }
Likely associated with the Be star HD\,199356 at a distance $\sim 700$ pc. Hence, possibly a gamma Cas star.
\subsection*{ SRGA J210501.3-353121 }
Likely associated with WISEA\,J210500.56-353141.7, $W1-W2\sim 1.0$, a radio source.
\subsection*{ SRGA J210906.7+480643 }
Likely associated with the star Gaia DR3 2165328756082487168 at a distance $\sim 1200$\,pc, which is an H$\alpha$-excess source \citep{2021MNRAS.505.1135F} and has been tentatively classified as a dwarf nova\footnote{ \url{https://www.aavso.org/vsx/index.php?view=detail.top&oid=686625}}.
\subsection*{ SRGA J211148.9+722812 }
A Seyfert 2 at $z=0.1061$ \citep{2023AstL...49...25U}.
\subsection*{ SRGA J211644.6+533408 }
An eclipsing CV with an orbital period of  0.2731 days \citep{2018AJ....155..247H}.
\subsection*{ SRGA J211703.9-485605 }
Associated with the galaxy 2MASX\,J21170265-4856128, $W1-W2\sim 0.3$.
\subsection*{ SRGA J211747.2+513848 }
A blazar at $z=0.0533$ \citep{2022ApJS..261....2K}.
\subsection*{ SRGA J212901.0+363110 }
Non-magnetic CV \citep{2023AJ....165..163C}.
\subsection*{ SRGA J213152.0+491400 }
Associated with the optical transient AT 2019weg = Gaia19fld = ZTF19acfixfe, a CV \citep{2021AJ....162...94S}.
\subsection*{ SRGA J214331.7+102854 }
Possibly associated with the star Gaia DR3 1765516439542838272, at a distance $\sim 1600$ pc.
\subsection*{ SRGA J214503.7+634527 }
Possibly associated with WISEA\,J214505.18+634541.2, $W1-W2\sim 1.3$, which is also a radio source. In the optical, confusion of several objects.
\subsection*{ SRGA J215706.5-694132 }
A broad-line radio galaxy \citep{2020A&A...641A.152A,2022ApJS..261....2K}.
\subsection*{ SRGA J220355.8+613222 }
Likely associated with the star Gaia DR3 2203373473312011008 at $\sim 400$\,pc, which is an H$\alpha$-excess source \citep{2021MNRAS.505.1135F} and has been classified as a CV based on variability\footnote{\url{https://www.aavso.org/vsx/index.php?view=detail.top&oid=844414}}.
\subsection*{ SRGA J220706.9+360935 }
Likely associated with WISEA\,J220707.91+360933.7, $W1-W2\sim 0.5$, also a radio source.
\subsection*{ SRGA J221913.5+362004 }
A Seyfert 2 at $z=0.14667$ \citep{2022AstL...48...87U}.
\subsection*{ SRGA J222516.9+012242 }
A gamma Cas star.
\subsection*{ SRGA J223132.7-750114 }
Possibly associated with the star Gaia DR3 6357827237727811712, at a distance $\sim1800$~pc.
\subsection*{ SRGA J223453.0-843445 }
A Seyfert 1 at $z=0.1630$ \citep{2022ApJS..261....5M}.
\subsection*{ SRGA J223715.8+402944 }
A Seyfert 1 at $z=0.0582$ \citep{2022AstL...48...87U}.
\subsection*{ SRGA J225413.1+690700 }
A CV \citep{2022A&A...661A..39Z}.
\subsection*{ SRGA J230338.2+465214 }
Associated with WISEA\,J230338.31+465159.5, a Seyfert 2 galaxy at $z=0.0460$ (Uskov et al., in prep.).
\subsection*{ SRGA J230631.0+155633 }
Associated with SDSS\,J230630.38+155620.4, a radio-loud Seyfert 2 galaxy at $z=0.439$ (Uskov et al., in prep.).
\subsection*{ SRGA J230641.7+550827 }
An intermediate polar \citep{2018AJ....155..247H}.
\subsection*{ SRGA J230821.7-782225 }
Likely associated with the galaxy IRAS\,23047-7838 = 6dFGS\,gJ230828.1-782223, $W1-W2\sim 1.4$, a radio source.
\subsection*{ SRGA J231951.5+564446 }
Possibly associated with WISEA\,J231950.48+564444.4, $W1-W2\sim 1.0$.
\subsection*{ SRGA J232038.1+482320 }
A Seyfert 2 at $z=0.04197$ \citep{2022AstL...48...87U}.
\subsection*{ SRGA J234620.9+812214 }
Possibly associated with the star Gaia DR3 2286390896172757760, at a distance $\sim 1.3$\,kpc.
\subsection*{ SRGA J235250.7-170448 }
A Seyfert 1 at $z=0.055$ \citep{2022AstL...48...87U,2022ApJS..261....2K}.
\subsection*{ SRGA J235335.2-454210 }
Likely associated with the galaxy LEDA\,523799, $W1-W2\sim 1.4$.

%% file: longtable_arxiv.tex
\onecolumn
\begin{landscape}
\label{tab:longtable}
\begin{longtable}{rlrrrcccrrrrrr}

\caption{Catalog of sources\tablefootmark{a)} detected during the first five surveys of the \srg/\art.}\\

\hline\hline
Id & Name & RA & Dec & R98 & S/N & Flux & Conventional name & Redshift & Class  \\
&  SRGA & (J2000)& (J2000) & ('') & & ($10^{-12}$~\flux)  & & &   \\
\hline
\endfirsthead
\caption{continued.}\\
\hline\hline
Id & Name & RA & Dec &  R98 & S/N & Flux & Conventional name & Redshift & Class  \\
&  SRGA & (J2000)& (J2000) & ('') & & ($10^{-12}$~\flux)  & & &   \\
\hline
\endhead
\hline
\endfoot
\input{tables/artxc_paper2_catalog_20240322-115324}
\end{longtable}
\tablefoot{ 
\tablefoottext{a}{For the description of columns see Section~\ref{s:catalog}. 
}
\tablefoottext{b}{Sources detected in X-rays for the first time by \srg/\art.}
\tablefoottext{c}{Additional information on identification and classification is provided in Section~\ref{s:notes}.}
}

\end{landscape}

%% file: tables/artxc_paper2_catalog_20240322-115324.tex
N001& J000048.1$-$070914&0.2005&$-$7.1539&20.5&6.36&5.69 $\pm$ 1.74&SWIFT J0001.0-0708&0.0375&SEYFERT$^{c)}$\\
N002& J000132.9+240237$^{b)}$&0.3869&24.0435&21.1&5.57&3.96 $\pm$ 1.36&&0.1045&SEYFERT$^{c)}$\\
N003& J000148.6$-$765739&0.4526&$-$76.9607&17.9&5.57&3.85 $\pm$ 1.22&SWIFT J0001.6-7701&0.0585&SEYFERT\\
N004& J000310.3+215744&0.7931&21.9622&21.5&5.81&3.58 $\pm$ 1.23&Mrk 334&0.0219&SEYFERT\\
N005& J000359.1+701922&0.9962&70.3227&18.7&6.21&2.19 $\pm$ 0.62&IGR J00040+7020&0.0960&SEYFERT\\
N006& J000549.7+102227&1.4571&10.3742&20.4&5.62&4.38 $\pm$ 1.47&RX J0005.8+1022&0.0948&SEYFERT\\
N007& J000632.6$-$690026&1.6357&$-$69.0071&15.4&8.17&5.63 $\pm$ 1.47&CTCV J0006-6900&&CV\\
N008& J000636.0+434219&1.6499&43.7052&19.7&6.67&3.61 $\pm$ 0.99&2RXS J000637.0+434223&0.1660&SEYFERT\\